\documentclass[%
 amsmath,amssymb,
 aps,
twocolumn,10pt,a4paper,sort&compress,pre,showpacs,superscriptaddress
]{revtex4-2}
\usepackage{graphicx} 
\usepackage{amsmath}
\usepackage{amssymb}
\usepackage{amsfonts}
\usepackage{bm}
\usepackage{dcolumn}

\usepackage{amsmath,amssymb,amsthm}
\usepackage{commath}
\usepackage{dcolumn}
\usepackage{hyperref}
\usepackage{etoolbox} 
\usepackage{siunitx}
\usepackage{multirow}

\usepackage{color}
\usepackage{xcolor}
\usepackage{overpic}

\usepackage[english]{babel}
\usepackage[caption=false]{subfig}
\addto\captionsenglish{}
\usepackage[capitalise,nameinlink]{cleveref}
\newcommand{\aref}[1]{\hyperref[#1]{Appendix}}
\usepackage{blkarray}

\usepackage[whole]{bxcjkjatype}
\begin{document}
\title{
Most probable path and invariant sets in noise-induced transition to turbulence
}

\author{Yoshiki Hiruta}
 \email{hiruta@rs.tus.ac.jp}
  \affiliation{
	Department of Physics and Astronomy, Faculty of Science and Technology, Tokyo University of Science, Chiba 278-8510, Japan}

\author{Kento Yasuda}
 \email{yasuda.kento@nihon-u.ac.jp}
  \affiliation{
	Laboratory of Physics, College of Science and Technology, Nihon University, Funabashi, Chiba 274-8501, Japan}
    
\author{Kenta Ishimoto}
 \email{kenta.ishimoto@math.kyoto-u.ac.jp}
 \affiliation{
	Department of Mathematics, Kyoto University, Kyoto 606-8502, Japan}
\date{\today}

\begin{abstract}
    Turbulence transition often arises from a subcritical transition between bistable states characterized by invariant sets of deterministic dynamical systems, and such transitions can be triggered by system noise as rare events. 
    In this study, we employ the Onsager-Machlup (OM) formulation of stochastic dynamics to examine the Hamilton equations governing the most probable transition paths (MPPs). We introduce an effective potential function, termed the OM potential, which depends on the noise strength. Focusing on the Dauchot-Manneville model as a minimal system with an edge state, we comprehensively analyze the MPP between laminar and turbulent states for different transition times. We find that the MPPs cross the separatrix at nearly the same point regardless of the transition time, and the obtained OM action values suggest that the transition to turbulence occurs more frequently than the transition to the laminar state. Moreover, we numerically demonstrate that the noise-induced transition paths follow the OM potential landscape and its bifurcation diagram, indicating that the qualitative behavior of the MPPs is determined by the OM potential. 
    Our methodology formulated in general dynamical systems
    provides a theoretical basis for predicting noise-induced transitions among invariant sets of the dynamics.
\end{abstract}

\maketitle

\section{introduction}

The transition from laminar to turbulent flow is one of the fundamental problems in fluid dynamics since Reynolds \cite{reynolds1883}. 
In a wide range of shear flows, 
turbulence arises through a subcritical transition, 
where sufficiently large disturbances develop into turbulence and infinitesimal disturbances decay through linear stability.
The subcritical nature of the transition also leads to bistability between laminar and turbulent states, 
observed in various fluid systems, including 
pipe flow \cite{avila2011,avila2013,Kerswell2005,shimizu2009driving,avila2023}, plane shear flow \cite{Nagata_1990,jimenez1991minimal,hamilton1995regeneration,itano_2001,kawahara_2001,gibson_2008,Marensi_2023,Engel2025}, and Kolmogorov flow \cite{hiruta_2015,hiruta_2017,hiruta_2020,hiruta2022}.

The presence of bistability in subcritical flows is typically characterized by the state space that contains 
two attractors separated 
by a basin boundary \cite{skufca2006, khapko2016edge}. 
The dynamical backbone of this boundary is 
the edge state, which is a saddle-type invariant solution whose invariant manifolds separate the laminar and turbulent basins \cite{itano_2001,kawahara_2001,kawahara_2012}. 
Since the stable manifold of the edge state forms a separatrix, the edge state has been a key structure in identifying the minimal seeds that trigger turbulence \cite{Pringle2010,Pringle2012,kerswell2018}. 
Moreover, states in the neighborhood of the edge state evolve along its unstable manifold, representing both turbulent bursting events and re-laminarization processes \cite{itano_2001,Tobias2007,wang2007,tobias2008}.

To capture the essential dynamical features of such subcritical transitions involving the edge state, 
low-dimensional ODE models based on Galerkin approximations have been proposed by Dauchot \& Manneville \cite{Dauchot1997} and Waleffe \cite{waleffe1995,waleffe1997}.  
These models successfully reproduce key characteristics in the full Navier–Stokes system, 
including the existence of edge states and energy-conservative nonlinear terms.
Among these reduced-order models, the Dauchot-Manneville model \cite{Dauchot1997}, which contains only two variables,  has been recently utilized as a minimal model in testing mathematical theories and hypothetical scenarios such as Lagrangian coherent structures \cite{beneitez2020edge}, bypass transition scenarios \cite{Beneitez2020}, reduced-order modeling via the spectrum submanifold (SSM) theory \cite{kaszas2023new} and flow control \cite{salmon2024closed}. 

Transition between metastable states has also been studied in stochastic models such as chemical reactions and gene-regulatory networks \cite{pollak2005reaction, tsimring2014noise, fang2019nonequilibrium}, where the transition is rather driven by intrinsic noise such as thermal noise. In a macroscopic phenomena such as fluid flow, it is not always tractable to incorporate all the detailed processes into a mathematical model, and noise terms are often added to the deterministic part. This additional noise may drive a stochastic transition between the metastable states as a rare event, which is called noise-induced transition \cite{forgoston2018primer}. Recently, metastable transitions in fluid systems have been investigated as a noise-induced transition, such as in developed turbulence \cite{xu2024} and circular reversals in large-scale flows \cite{podvin2017precursor, Soons2025},
as well as the transition to turbulence \cite{bouchet2019rare, gome2022extreme}.

Wang et al. \cite{Wan_2015} analyzed a two-dimensional Poiseuille flow under an infinitesimally small external noise with a stochastic forcing term. They numerically detected the rare transition as a transition path by minimum action algorithm \cite{WAN2013} using the Freidlin-Wentzel theory \cite{Fredlin1984}.
They proposed a scenario that a transition path first travels towards the vicinity of a {\it pseudo transition state} located on the separatrix and follows the separatrix towards the edge state. More recently, Rolland \cite{rolland2024} studied a plane Couette flow by sampling rare events \cite{Rolland_2022} and numerically detected the transition path, suggesting that the transition path 
visits a {\it mediator point} on the separatrix before reaching the vicinity of the edge state at a small noise limit, consistent with the conjecture proposed by Ref.~\cite{Wan_2015}. Further, Rolland \cite{rolland2024} proposed another transition scenario at a finite noise strength that transition paths should cross the separatrix away from the edge state.

These studies provide a connection between dynamic information of deterministic systems and the typical transition dynamics of stochastic systems. However, it has been numerically challenging to systematically specify the transition path, because of the high dimensionality of the Navier–Stokes equations, the underlying chaotic nature of turbulence, and the rarity of transition events under weak noise. Moreover, the transition events often contain several transition time scales, which further limits the reproducibility of the transition path. Hence, to comprehensively address the universal feature of transition path in the noise-induced turbulence, one requires a systematic methodology that connects the deterministic and stochastic system.

Recently, as a powerful framework 
to investigate stochastic transition, the most probable path (MPP) has increasingly been adopted in various research areas including climate models, gene-regulatory networks, excitable systems and active soft matter \cite{li2021, hu2022onsager, yasuda2022, Borner2024, yasuda2025demon}, since theoretical outcomes have been experimentally validated
\cite{gladrow2021experimental}.
The MPP is formulated as a minimizer of the Onsager-Machlup (OM) action defined as the path integral of stochastic dynamics \cite{Onsager1953,risken1989}
and is expected to be realized in a small but finite noise intensity. Notably, at finite noise strength, the OM formulation significantly deviates from the Freidlin-Wentzel theory \cite{gladrow2021experimental,Borner2024}.

In this study, based on the OM formulation, 
we aim to provide a unified picture of typical transition paths in noise-induced turbulence transitions under finite noise intensity, by clarifying the relation between invariant sets of a deterministic model and the most probable path of the associated stochastic model.
To address universal properties of the most probable path in a subcritical transition to turbulence, we present the Hamiltonian structure of the the OM formulation and introduce an effective potential function, which we refer to as the OM potential, demonstrating that the potential function determines qualitative behavior of the transition path.
To illustrate the usability of our methodology, in this paper, we focus on a stochastic version of the Dauchot-Manneville (DM) model as a minimal system with an edge state and comprehensively analyze the most probable path for various transition times and noise intensities, while the methodology presented here is straightforwardly applicable to higher dimensional models.

This paper is organized as follows.
In Sec.~\ref{sec:setting}, we formulate the Onsager–Machlup action for a general dynamical system and present the Hamiltonian structure of the most probable path, together with our numerical schemes. After introducing the Dauchot–Manneville model in Sec.~\ref{sec:modeldef}, we present the similarity between the probability density function and the potential function of the Hamiltonian equation in Sec.~\ref{sec:density}.
Section \ref{sec:mpp} is devoted to numerical results of the most probable path, with a particular focus on the transition between the laminar and turbulent states.
Finally, with the dynamical system's description of the MPP, we show how a simple invariant solution governs the noise-induced transition between laminar and turbulent states in Sec.~\ref{sec:dynsys}. Section \ref{sec:conclusion} is concluding remarks.
       
\section{General Formulation and Numerical schemes\label{sec:setting}}
\subsection{Deterministic and Stochastic Systems}

We consider a $N$-dimensional stochastic system for a state variable $\bm{x}\in \mathbb{R}^{N}$,
as a sum of an autonomous deterministic process $d\bm{x}/dt=\bm{f}(\bm{x})$ and a random process, given by the following
Langevin equation with noise intensity $D$ \footnote{The noise intensity $D$ may differ from the standard definition of the diffusion constant by a factor of $2$. \cite{risken1989}.}
\begin{align}
\frac{d\bm{x}}{dt} &= \bm{f}(\bm{x}) +\sqrt{D}\,\bm{\xi}(t), \label{eq:langevin}
\end{align}
where $\bm{\xi}(t)$ is the zero-mean white Gaussian noise satisfying $\langle \bm{\xi}\bm{\xi}^{\textrm{T}}\rangle=I$, with the $N$-dimensional identity matrix $I$.
With a discrete representation $\Delta\bm{x} \equiv \bm{x}(t+\Delta t)-\bm{x}(t)$ for a short-time duration $\Delta t$, the stochastic equation Eq.~\eqref{eq:langevin} is then rewritten by the Stratonovich interpretation as 
\begin{align}
\Delta\bm{x} &=\frac{1}{2}\left[\bm{f}(\bm{x}(t+\Delta t))+\bm{f}(\bm{x}(t))\right] +\Delta \bm{W}, \label{eq:shorttime}
\end{align}
where $\Delta \bm{W}$ is the associated Wiener process, with its probability density 
being proportional to $\exp(-|\Delta \bm{W}|^2/(2D\Delta t))$.
Each realization of $\Delta \bm{W}$, hence that of $\bm{x}$, is unbounded, leading to a nonzero transition probability between any two arbitrary states during the time duration $\Delta t$. Such a transition may occur even when the dynamical system $d\bm{x}/dt=\bm{f}(\bm{x})$ 
possesses a separatrix that divides the phase space into different basins of attractors.

\subsection{Onsager--Machlup Action}

We now evaluate the probability of a path of the state $\{\bm{x}\}_{t=0}^{\tau}$ from time $t=0$ to a certain finite time at $t=\tau$, using the Onsager-Machlup (OM) action $S$ \cite{Onsager1953}.
The transition probability
$P(\bm{X}_{\mathrm{init}},\bm{X}_{\mathrm{final}})$ between 
the initial state $\bm{x}(0)=\bm{X}_{\mathrm{init}}$ and 
the final state $\bm{x}(\tau)=\bm{X}_{\mathrm{final}}$ is obtained via the path integral formula as 
\begin{align}
P(\bm{X}_{\mathrm{init}},\bm{X}_{\mathrm{final}})\propto 
\int \mathcal{D}\bm{x} \exp\left(-\frac{1}{D}S (\{\bm{x}\}_{t=0}^{\tau}) \right),
\end{align}
where the integration $\int \mathcal{D}\bm{x}$ is conducted over all paths $\{\bm{x}\}_{t=0}^{\tau}$ 
between the initial and final states.
Here, the OM action $S$  for the path $\{\bm{x}\}_{t=0}^{\tau}$ is defined as 
\begin{align}
S (\{\bm{x}\}_{t=0}^{\tau}) \equiv \int_0^\tau  \mathcal{L} (\bm{x},\dot{\bm{x}})dt,   
\end{align}
where $ \mathcal{L}$ is the associated OM Lagrangian, given by 
\begin{align}
\mathcal{L} \equiv\frac{1}{2} \left(\dot{\bm{x}}-\bm{f}(\bm{x})\right)^{\textrm{T}}\left(\dot{\bm{x}}-\bm{f}(\bm{x})\right)+\frac{D}{2} \bm{\nabla} \cdot \bm{f},    
\end{align}
with the superscript symbol $\textrm{T}$ denoting the matrix transpose.

The transition path that minimizes the action $S_{\mathrm{OM}}$ exhibits the highest probability among the possible paths with the initial and final states, hence 
it is called the most probable path (MPP). In the weak noise limit, the transition probability is well represented solely by the most probable path via the saddle-point approximation and is regarded as an instanton \cite{Grafke2015} as an analogue to the path integral formulation of quantum field theory.

\subsection{Most probable path \label{sec:mppkinematics}}
The minimizer of the OM action $S_{\mathrm{OM}}$ or equivalently the MPP follows the Euler-Lagrange (EL) equations,
\begin{align}
    \frac{d\bm{x}}{dt}&=\bm{f}(\bm{x})+\bm{p},\label{eq:EL1}\\ 
    \frac{d\bm{p}}{dt}&=-J^{\textrm{T}}(\bm{x}) \bm{p}+\frac{D}{2} \bm{\nabla} (\bm{\nabla}\cdot\bm{f}(\bm{x})). \label{eq:EL2} 
\end{align}
Here, $\bm{p}$ is the generalized momentum associated with the OM Lagrangian $\mathcal{L}$, defined as $\bm{p}= \partial \mathcal{L}/\partial \dot{\bm{x}}$, and $J(\bm{x})$ is the Jacobian matrix of $\bm{f}(\bm{x})$.

 By eliminating the generalized momentum from the EL equations \eqref{eq:EL1}-\eqref{eq:EL2}, we obtain the same equation as that of a particle motion in an electromagnetic field, given by 
\begin{align}
\ddot{\bm{x}}&=\bm{E}(\bm{x})+B(\bm{x})\bm{\dot{x}} \label{eq:ELEM0},
\end{align}
with the electric field $\bm{E}(\bm{x})\equiv-\bm{\nabla}V(\bm{x})$ and 
the antisymmetric tensor associated with the magnetic field $B(\bm{x})\equiv J(\bm{x})-J^{\mathrm{T}}(\bm{x})$~\cite{wang2010}.
The effective electrostatic potential $V$ is determined by the deterministic dynamics $\bm{f}$ and the noise intensity $D$ as
\begin{align}
V(\bm{x})&\equiv-\frac{|\bm{f}(\bm{x})|^2}{2}-\frac{D}{2}\bm{\nabla}\cdot\bm{f}(\bm{x})\label{eq:potential0},
\end{align}
we hereafter refer to as {\it Onsager-Muchlup potential}, or simply OM potential. 
One can readily find that a fixed point
of the EL equations $\bm{X}^{\ast}$ 
is a local minimum or maximum of $V$ 
and $\bm{\nabla}V(\bm{X}^{\ast})=0$.

Interestingly, the OM potential function is found to be identical to the quantity known as the effective potential in the context of the Fokker-Planck equation~\cite{risken1989}, where the gradient system in the form of $\bm{f}=-\bm{\nabla}U$ is usually assumed [See Appendix \ref{sec:schroedinger}]. 
We note that the OM potential $V(\bm{x})$ here is distinct from the potential function $U(\bm{x})$ of the gradient system. 
In the gradient system, the stationary probability density function is proportional to $\exp(-U/D)$, whereas such a relation does not hold for $V$. Further, an analytical expression of the probability density function is, in general, not available. See Appendix \ref{sec:schroedinger} for further discussion.

With the Legendre transformation, we introduce the OM Hamiltonian as $\mathcal{H} \equiv \bm{p}\cdot \dot{\bm{x}}-\mathcal{L}$ \cite{li2021}. Hence the total energy,
\begin{align}
    E=K(\dot{\bm{x}})+V(\bm{x}), \label{eq:energy}
\end{align}
is a constant over time evolution, where the kinetic energy is given by $K\equiv |\dot{\bm{x}}|^2/2$.
The OM action is then rewritten as
\begin{align}
    S &= Q -E\tau,
    \label{eq:thermodynamics}
\end{align}
with 
\begin{align}
    Q&\equiv\int_0^{\tau} \bm{p}\cdot\dot{\bm{x}}dt.
\end{align}

The quantity $Q$ may be physically interpreted as the work acting on the system by the heat reservoir. Hence, we here call $Q$ the heat absorption. 
When the work is generated from the heat reservoir to the system ($\bm{p}\cdot\dot{\bm{x}}>0$), this work contributes to increasing the OM action. 
We note that $Q$ depends only on the shape of the path and does not depend on the rate of the process.

Following Ref.~\cite{gladrow2021}, we introduce the relative path probability between a pair of arbitrary paths, $\{\bm{x}_1\}$ and $\{\bm{x}_2\}$ sharing the same initial and final times $[0, \tau]$, where their initial and final states may be taken arbitrarily. 
We first define the sojourn probability $P^{\{\bm{x}\}}_R$ as the probability that a stochastic trajectory remains within a moving ball of radius $R$ around a path $\{\bm{x}\}_{t=0}^\tau$. The relative path probability is obtained by the ratio of the two sojourn probabilities in the limit of $R\rightarrow 0$ as
\begin{align}
    \frac{e^{-S(\{\bm{x}_1\})}}{e^{-S(\{\bm{x}_2\})}}=\lim_{R\rightarrow 0} \frac{P^{\{\bm{x}_1\}}_R}{P^{\{\bm{x}_2\}}_R}
    \label{relativeprop},
\end{align}
which is simply $\exp(-\Delta S)$ with $\Delta S=S(\{\bm{x}_1\})-S(\{\bm{x}_2\})$.

We then consider the relative path probability between MPPs with different initial and final states. By denoting
the solutions of the EL equations as $\{\bm{x}_1, \bm{p}_1\}$ and $\{\bm{x}_1, \bm{p}_1\}$, respectively, 
we may write the difference of the action values between the two paths as
\begin{align}
    \Delta S &= \int_0^{\tau} (\bm{p}_1\cdot\dot{\bm{x}}_1-\bm{p}_2\cdot\dot{\bm{x}}_2)dt -(E_1-E_2)\tau \\
    &=\Delta Q-\Delta E\cdot\tau\label{eq:dS}.
\end{align}
Here, $\Delta Q=Q_1-Q_2$ and $\Delta E=E_1-E_2$ represent the differences in heat absorption and energy of the two MPPs.

 \subsection{Numerical methods}
 \label{sec:numeric}
To obtain the MPP from the OM action, we employ the shooting method to numerically solve the boundary-value problem of the EL equations \eqref{eq:EL1}-\eqref{eq:EL2} with the boundary conditions, $\bm{x}(0) =\bm{X}_{\textrm{init}}$ and $\bm{x}(\tau) =\bm{X}_{\textrm{final}}$. 

Here we introduce a flow map $G_{t}: \mathbb{R}^{2N}\to \mathbb{R}^{2N}$ for the pair of position and momentum, $\bm{w}=(\bm{x},\bm{p})$, to represent the time evolution of the EL equations, \eqref{eq:EL1}-\eqref{eq:EL2}, as
\begin{align}
     \bm{w}(t_0+t)&\equiv\bm{G}_t(\bm{w}(t_0))
\end{align}
for an arbitrary time $t_0$ and time duration $t$.
The flow map $\bm{G}_{t}$ is numerically constructed by the 4th-order Runge-Kutta method for ordinary differential equations with a sufficiently small time discretization to ensure the energy conservation as  required by the Hamiltonian structure of the system.
Our task is, therefore, to find a pair $(\bm{p}_{\mathrm{init}}, \bm{p}_{\mathrm{final}})$  that satisfies 
\begin{align}
    \bm{G}_\tau((\bm{X}_{\mathrm {init}},\bm{p}_{\mathrm{init}}))-(\bm{X}_{\mathrm{final}},\bm{p}_{\mathrm{final}})=\bm{0}.
\end{align} 
This problem, however, becomes computationally challenging for a large $\tau$ due to a bad condition number.  
To overcome this difficulty, we employ the multiple shooting method. We divide the time interval $[0,\tau]$ into equally-partitioned $M$ small intervals 
  \begin{align}
      0=\tau_0<\tau_1<\cdots<\tau_M=\tau,
  \end{align}
with $\tau_m=m\tau/M$ ($m=0,1,\dots,M$). By applying the formula $G_{\tau_1}G_{\tau_2}=G_{\tau_1+\tau_2}$, we may rewrite the boundary-value problem by $M+1$ equations,
\begin{align}
      \bm{G}_{\delta\tau_0}((\bm{X}_{\mathrm{init}},\bm{p}_{\mathrm{init}}))&=\bm{w}_{1} \label{eq:G1} \\
      \bm{G}_{\delta\tau_m}(\bm{w}_m)&=\bm{w}_{m+1} ~(m=1,\cdots,M-1)\\
      \bm{G}_{\delta\tau_{M-1}}(\bm{w}_{M-1})&=(\bm{X}_{\mathrm{final}},\bm{p}_\mathrm{final} ) \label{eq:G3},
\end{align}
with $\delta \tau_m=\tau_{m+1}-\tau_{m}=1/M$.
The problem is then reduced to finding zeros of a $2NM$-dimensional nonlinear map $\bm{H}_{\bm{X}_{\mathrm{init}};\bm{X}_{\mathrm{final}}}:\mathbb{R}^{2NM}\to\mathbb{R}^{2NM}$, given by
  \begin{align}
  \bm{H}_{\bm{X}_{\mathrm{init}};\bm{X}_{\mathrm{final}}}(\bm{p}_{\mathrm{init}},\bm{p}_\mathrm{final},\bm{w}_1,\cdots\bm{w}_{M-1}) \nonumber
  \\=\begin{pmatrix}
      \bm{G}_{\delta\tau_0}((\bm{X}_{\mathrm{init}},\bm{p}_{\mathrm{init}}))-\bm{w}_{1}  \\
      \bm{G}_{\delta\tau_1}(\bm{w}_1)-\bm{w}_2\\
    \vdots\\
      \bm{G}_{\delta\tau_{M-2}}(\bm{w}_{M-2})-\bm{w}_{M-1}\\
      \bm{G}_{\delta\tau_{M-1}}(\bm{w}_{M-1})-(\bm{X}_{\mathrm{final}},\bm{p}_\mathrm{final} )
  \end{pmatrix}\label{eq:Gz3}.
  \end{align}
We numerically construct the flow map \label{eq:Gz3} and use the Newton-GMRES method to obtain $2NM$ unknowns $(\bm{p}_{\mathrm{init}},\bm{p}_\mathrm{final},\bm{w}_1,\cdots\bm{w}_{M-1})$. 

The convergence of the Newton method strongly depends on the initial guess. To explain our choice of the initial data of the scheme, we first introduce rescaled variables, $\tilde{t}=t/\tau$, $\tilde{\bm{x}}=\bm{x}$, and $\tilde{\bm{p}}=\tau \bm{p}$, leading to the rescaled EL equations,
\begin{align}
  \dot{\bm{\tilde{x}}}&=\tau \bm{f}+\dot{\tilde{\bm{p}}}, \\
  \dot{\bm{\tilde{p}}}&=-\tau J^{T}\tilde{\bm{p}}+D\tau^2\bm{\nabla} (\bm{\nabla}\cdot\bm{f})/2, 
\end{align}
where the dot symbol denotes the derivative with respect to the rescaled time $\tilde{t}$.
The rescaled equations imply that the constant momentum 
\begin{align}
(\tilde{\bm{x}},\tilde{\bm{p}})=\left(\frac{(\bm{X}_{\mathrm {final}}-\bm{X}_{\mathrm {init}})t}{\tau}+\bm{X}_{\mathrm {init}},\frac{\bm{X}_{\mathrm {final}}-\bm{X}_{\mathrm {init}}}{\tau}\right)
\label{eq:constmomentum}
\end{align}
provides a good approximation of the EL solution if $\tau\ll1$, and notably, Eq.~\eqref{eq:constmomentum} is independent of $\bm{f}$. We use this constant momentum data as our initial guess 
of the numerical scheme for small $\tau$ and the numerical continuation method 
is used to trace a branch of the solutions for higher $\tau$. 

Nonetheless, the EL equations only ensure that the solution is an extremum and its solution is not necessarily a global minimum of the action. To guarantee the global property of the obtained solutions, we independently implemented a neural network architecture to numerically approximate the MPP by directly minimizing the OM action. We then confirmed that the solutions from both methods are in good agreement, 
and we thus conclude that the MPP obtained by shooting method is actually a global minimum of the OM action. In the results sections below, we only show the MPP obtained by the shooting method for better numerical precision. Details of the neural network architecture are explained in Appendix \ref{sec:neural}.

\begin{figure*}
    \centering
    \begin{overpic}[width=0.49\linewidth]{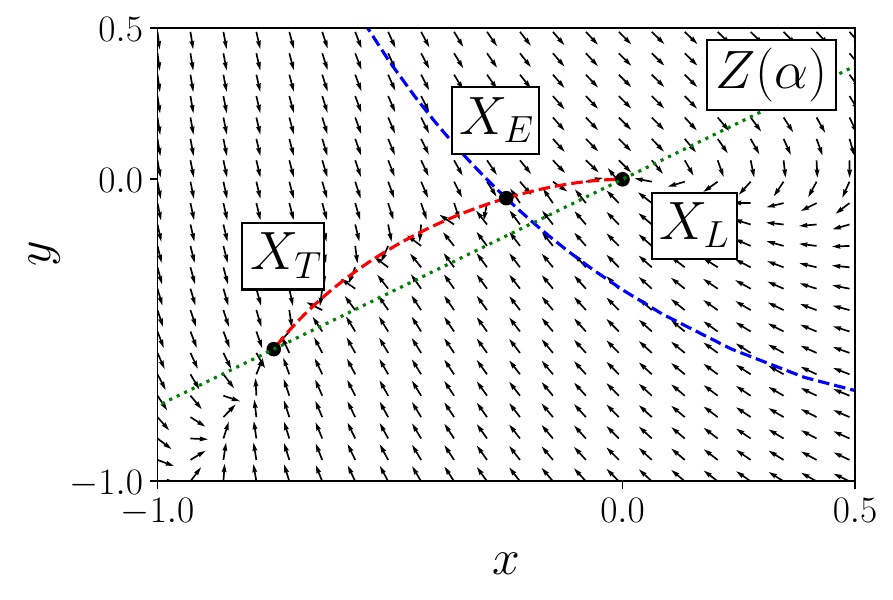}
    \put(3,60){(a)}
    \end{overpic}\begin{overpic}
    [width=0.49\linewidth]{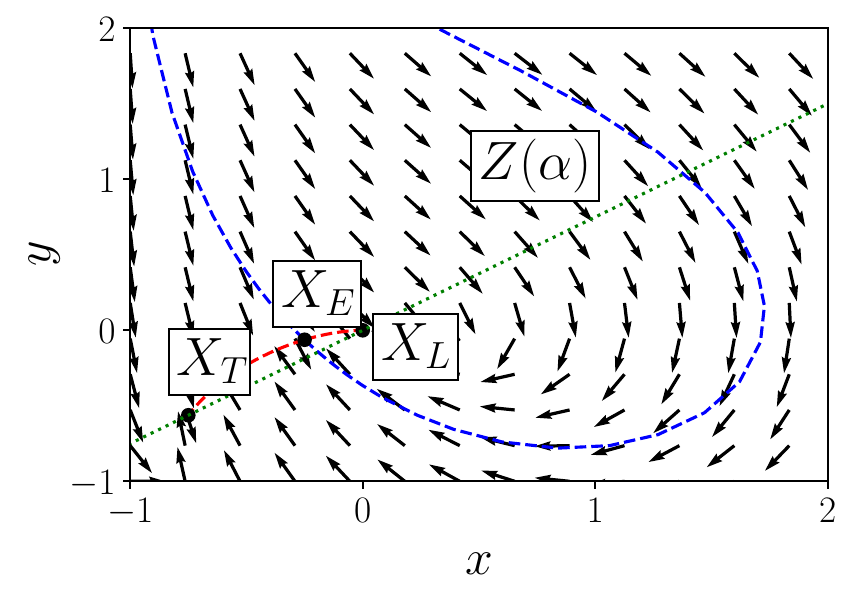}
    \put(3,60){(b)}
    \end{overpic}
    \caption{ Invariant manifolds and flow of the DM model with $s_1=-0.1875$ and $s_2=-1$ in  
    (a) narrow and (b) broad region of the $(x,y)$-plane. 
    The saddle point $\bm{X}_E$ contains two unstable manifolds [red dashed lines], which are heteroclinic orbits connecting the two 
    stable points, $\bm{X}_L$ and $\bm{X}_T$. The stable manifolds of the saddle points [blue dashed line] is a separatrix that divides the basins of attractions for the two stable points.
    Arrows indicate the direction of the flow [Eqs.~\eqref{eq:DM1} and \eqref{eq:DM2}] and 
    the green dotted line is the straight line $\bm{Z}(\alpha)$ [Eq.~\eqref{eq:Z}] that passes through the two stable points, $\bm{X}_{L}$ and $\bm{X}_T$. 
    } 
    \label{fig:phase}
\end{figure*}

 \section{model definition\label{sec:modeldef}}
 \subsection{The Dauchot--Manneville model}
 We consider a two-dimensional ODE model of subcritical transition introduced by 
 Dauchot \& Manneville \cite{Dauchot1997,Beneitez2020}. 
While the Dauchot-Manneville (DM) model was proposed as a conceptual mathematical model, this simple ODE system retains some of essential physics of the subcritical transition in a shear flow. Explicitly, the DM model is given by
\begin{align}
    \frac{dx}{dt}&= s_1 x +y+xy \label{eq:DM1},\\
    \frac{dy}{dt}&= s_2 y -x^2 \label{eq:DM2},
\end{align}
where the variables $\bm{x}=(x,y)$ physically represent dominating modes of the velocity deviation fields from the base flow, and the two parameters $s_1<0$ and $s_2<0$ indicate viscous effects. The nonlinear terms $xy$ and $-x^2$ originate from inter-mode coupling, 
and the second term $y$ on the right-hand side of Eq.~\eqref{eq:DM1} represents the coupling with the shearing base flow. 

The kinetic energy of the DM model, defined as $E_{\mathrm{DM}}\equiv(\dot{x}^2+\dot{y}^2)/2$, follows the time evolution as
    $dE_{\mathrm{DM}}/dt= s_1 x^2+s_2 y^2 +xy$.
Contributions from the nonlinear terms are all canceled, as is always the case with the Navier-Stokes equations,
and the dissipation $s_1 x^2+s_2 y^2\leq0$ is balanced with energy transfer from
the base flow $xy$.

when $\Delta\equiv1-4s_1s_2>0$ is satisfied, the DM model possesses three fixed points, given by
\begin{align}
\bm{X}_L&=(0,0),\\
\bm{X}_E&=\left(\frac{-1+\sqrt{\Delta}}{2},\frac{(-1+\sqrt{\Delta})^2}{4s_2}\right),\\
\bm{X}_T&=\left(\frac{-1-\sqrt{\Delta}}{2},\frac{(-1-\sqrt{\Delta})^2}{4s_2}\right).
\end{align}
Here the fixed point $\bm{X}_L$ represents the laminar solution and is always a solution to DM model. The two additional fixed points, $\bm{X}_E$ and $\bm{X}_T$, emerge via a saddle node bifurcation at $\Delta=0$ as counterparts of the edge state and turbulent state of the subcritical transition system, respectively.

In Fig.~\ref{fig:phase}, the flow dynamics in the $(x,y)$-plane with $s_1=-0.1875$ and $s_2=-1$ are shown. This parameter set is the same setting as in Ref.~\cite{Beneitez2020} and is used throughout this paper.
When $\Delta>0$, as seen in Fig.\ref{fig:phase}, the basins of attraction for the two stable fixed points, $\bm{X}_L$ and $\bm{X}_T$,
are divided by a stable manifold of $\bm{X}_E$ [blue dashed line in Fig.~\ref{fig:phase}], and connected by an unstable manifold of $\bm{X}_E$ [red dashed line in Fig.\ref{fig:phase}]. Hence, the unstable manifold is a heteroclinic orbit of the DM model.

To further quantify the flow structure of the model, we parameterize the heteroclinic orbit by normalized arclength $\tilde{s}\in[0,1]$ such that
$\tilde{\bm{X}}(\tilde{s})=(\tilde{X}(\tilde{s}),\tilde{Y}(\tilde{s}))$,
with $\tilde{\bm{X}}(0)=\bm{X}_L$ and $\tilde{\bm{X}}(1)=\bm{X}_T$ as well as the edge state 
$\tilde{\bm{X}}(\tilde{s})=\bm{X}_E$ with $\tilde{s}\approx0.265$. We numerically construct this parameterization by the cubic spline method.

For later use, we introduce a straight line that passes $\bm{X}_{L}$ and $\bm{X}_T$, which is parameterized by $\alpha\in\mathbb{R}$ [the green dotted line in Fig.~\ref{fig:phase}] such that
\begin{align}
\bm{Z}(\alpha)=(1-\alpha)\bm{X}_{L}+\alpha\bm{X}_{T}.  \label{eq:Z}
\end{align}
We find that $\bm{Z}(\alpha)$ has two intersections with the separatrix at $\alpha\approx 0.258$ and $\alpha\approx -1.84$.

\begin{figure*}
    \centering
    \begin{overpic}[width=0.49\linewidth]{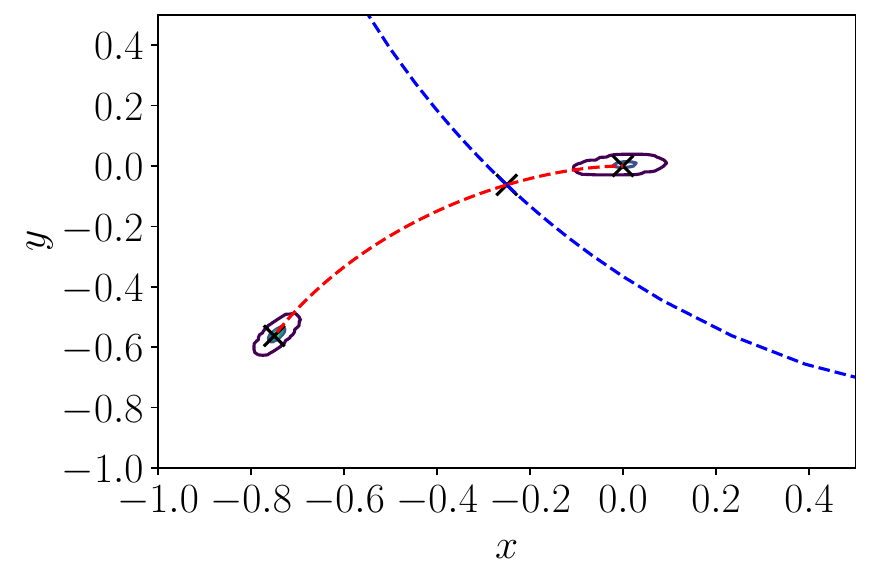}
    \put(3,60){(a)}
    \put(24,32){$\bm{X}_T$}
    \put(55,41){$\bm{X}_E$}
    \put(67,51){$\bm{X}_L$}
    \put(80,57){$D=10^{-4}$}
    \end{overpic}\begin{overpic}
    [width=0.49\linewidth]{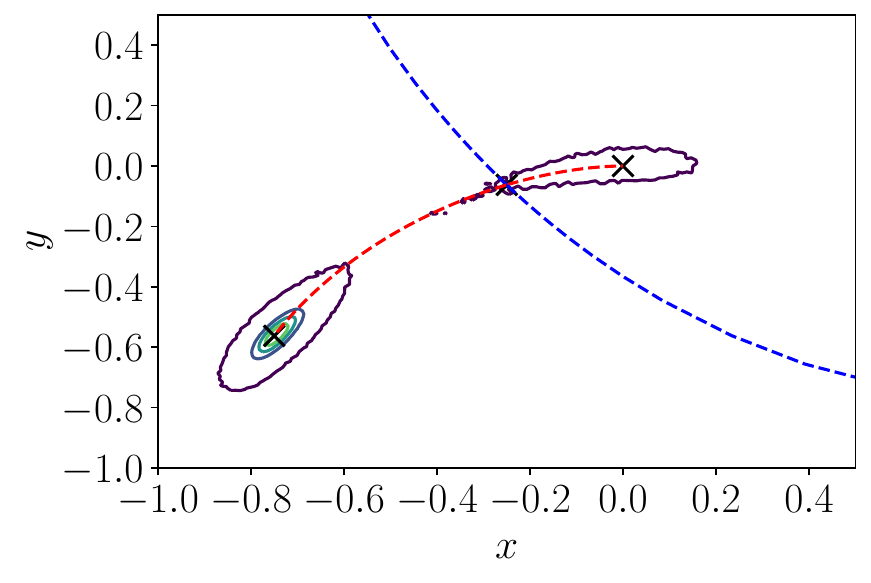}
    \put(3,60){(b)}
    \put(24,32){$\bm{X}_T$}
    \put(55,41){$\bm{X}_E$}
    \put(69,51){$\bm{X}_L$}
    \put(80,57){$D=10^{-3}$}
    \end{overpic}\\ \begin{overpic}
    [width=0.49\linewidth]{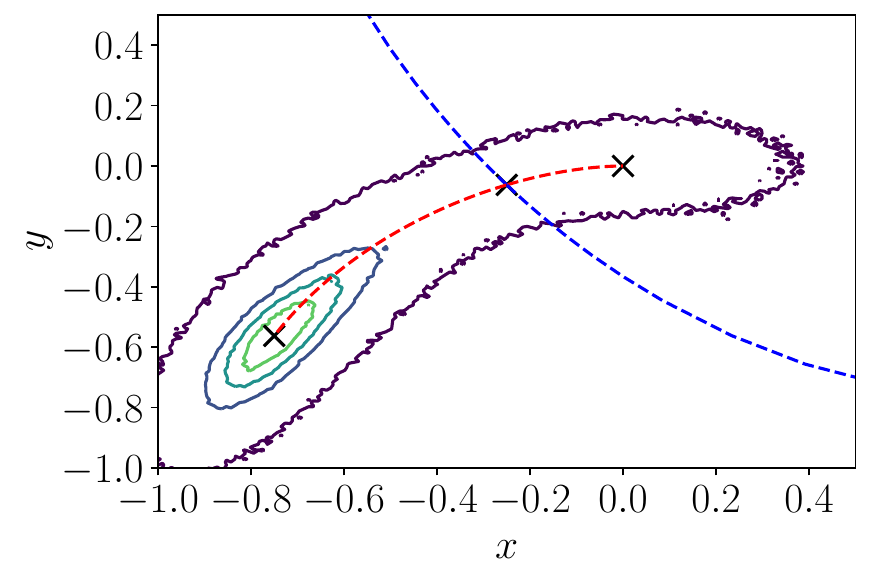}
    \put(3,60){(c)} 
    \put(80,57){$D=10^{-2}$}
    \end{overpic}\begin{overpic}
    [width=0.49\linewidth]{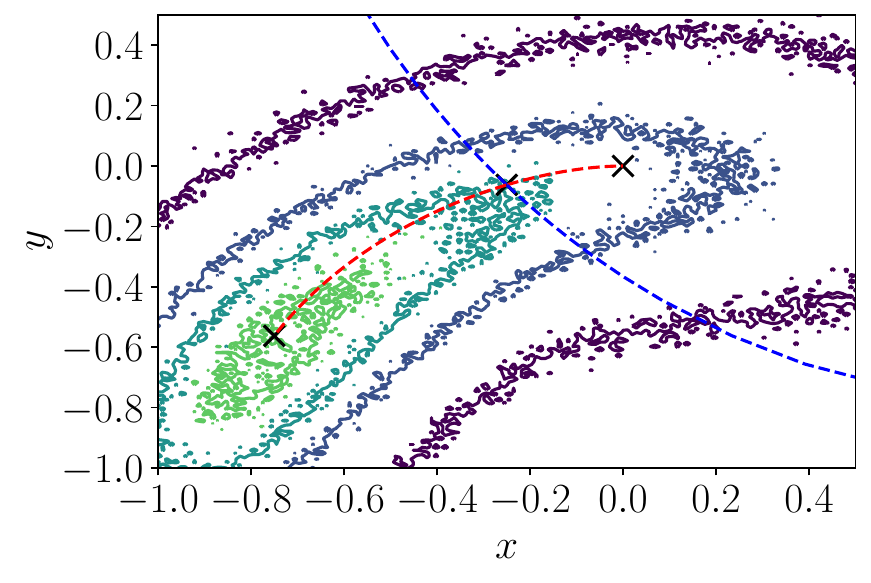}
    \put(3,60){(d)}
    \put(80,55){$D=10^{-1}$}
    \end{overpic}
    \caption{Probability density distribution (PDF) of the Langevin simulation of the stochastic DM model, $P(x,y)$, for different noise strengths with (a) $D=10^{-4}$, (b) $D=10^{-3}$, (c) $D=10^{-2}$, (d) $D=10^{-1}$. The maximum value of $P(x,y)$ is normalized to unity and contours of the PDF (isolines of $P=0, 0.25, \dots, 1$) are shown in color. The three fixed points, together with the stable and unstable manifolds of the edge state, are shown as in Fig.~\ref{fig:phase}. 
    }
    \label{fig:pdf}
\end{figure*}

\begin{figure}
    \centering
    \begin{overpic}[width=\linewidth]{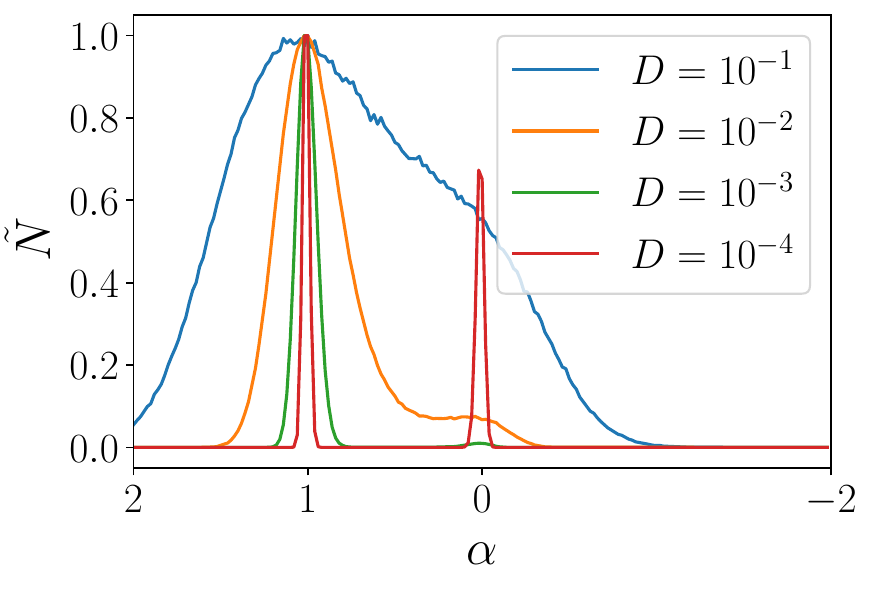}
    \put(37,10){$\nwarrow$}
    \put(58,10){$\nwarrow$}
    \put(42,7){$\bm{X}_T$}
    \put(63,7){$\bm{X}_L$}
    \end{overpic}
    \caption{Normalized marginal PDF $\tilde{N}(\alpha)$ along  the $\bm{Z}(\alpha)$ line, with $D=10^{-1}$, $10^{-2}$, $10^{-3}$, and $10^{-4}$. The horizontal axis is flipped so that relative position matches with the plots in Fig.~\ref{fig:phase} and Fig.~\ref{fig:pdf}, where the left and right peaks at $\alpha=1$ and $\alpha=0$ correspond to $\bm{X}_T$ and $\bm{X}_L$, respectively. 
    \label{fig:peaks}}
\end{figure}
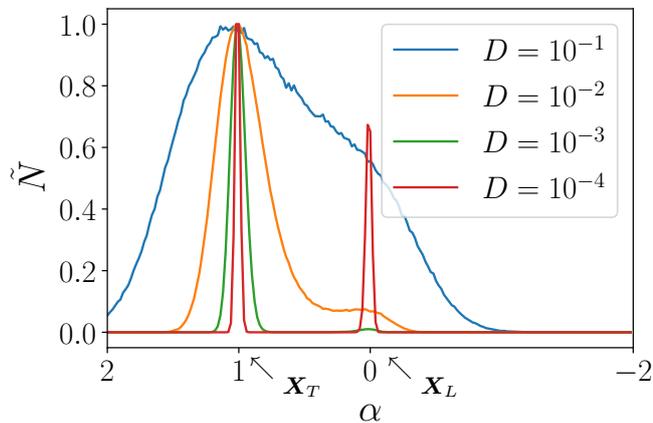
 \subsection{Euler--Lagrange equation for DM model}
 The most probable path of the stochastic DM model is obtained by the EL equations,
\begin{align}
    \frac{dx}{dt}&= s_1 x +y+xy+p_x ,\label{eq:ELDM1}\\
    \frac{dy}{dt}&= s_2 y -x^2+p_y,\\
    \frac{dp_x}{dt}&= -(s_1+y) x +2xp_y,\\
    \frac{dp_y}{dt}&= -(x+1)p_x -s_2p_y+\frac{D}{2}.\label{eq:ELDM4},
\end{align}
where the effect of noise intensity $D$ only appears on the right-hand-side of Eq.~\eqref{eq:ELDM4}.
Although the degrees of freedom in the EL equations are doubled from the original DM system \eqref{eq:DM1}-\eqref{eq:DM2}, we may reduce the EL equations to the two-dimensional deterministic system by using the analogy to the motions in an electromagnetic field [see Eq.\eqref{eq:ELEM0}], as
\begin{align}
\Ddot{\bm{x}}=\bm{E}+\begin{pmatrix}
  0 & (3x+1) \\
    -(3x+1) &0
\end{pmatrix}\dot{\bm{x}} \label{eq:ELEM},
\end{align}
where the corresponding electric potential, $\bm{E}=-\nabla V$, is obtained as
\begin{align}
V=-\frac{1}{2}\left[(s_1x+y+xy)^2+(s_2y-x^2)^2+D(s_1+s_2+y)\right]\label{eq:potential},
\end{align}
which is the OM potential function of the DM model.
The magnetic field is applied normal to the $(x,y)$-plane with its strength $3x+1$. The rotational motion associated with this magnetic field is therefore counter-clockwise when $x>-1/3$ and is reversed to clockwise when $x<-1/3$.

 \section{probability Density and Onsager-Muchlup potential landscape \label{sec:density}}

Before moving to a detailed analysis of the MPP solution, in this section we examine static features of the stochastic DM model, focusing on the stationary probability distribution function (PDF), and discuss its relation to the OM potential landscape.

\begin{figure*}
    \centering
    \begin{overpic}[width=0.45\linewidth]{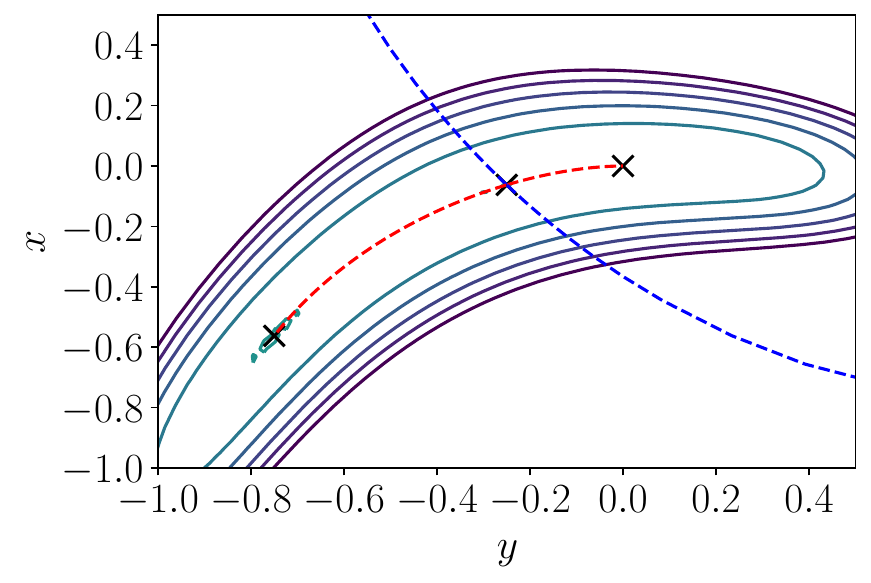}
    \put(3,60){(a)}
    \put(25,24){$\bm{X}_T$}
    \put(54,41){$\bm{X}_E$}
    \put(73,49){$\bm{X}_L$}
    \put(80,60){$D=10^{-3}$}
    \end{overpic}\begin{overpic}
    [width=0.45\linewidth]{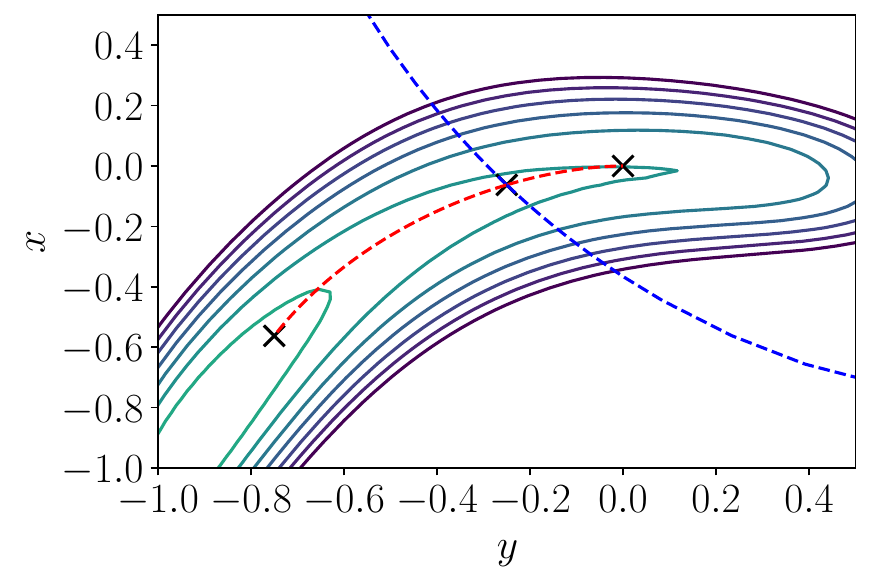}
    \put(3,60){(b)}
    \put(25,24){$\bm{X}_T$}
    \put(54,41){$\bm{X}_E$}
    \put(73,49){$\bm{X}_L$}
    \put(80,60){$D=10^{-1}$}
    \end{overpic}
    \caption{
    Contours of the OM potential function $V(x,y)$ [Eq.~\eqref{eq:potential}] with (a) $D=10^{-3}$ and (b) $D=10^{-1}$. The three fixed points, together with the stable and unstable manifolds of the edge state, are shown as in Figs.~\ref{fig:phase} and \ref{fig:pdf}. Isolines of $V=-0.1, 0.012,\dots, 0.1$ are shown in color.
    }
    \label{fig:potential}
\end{figure*}

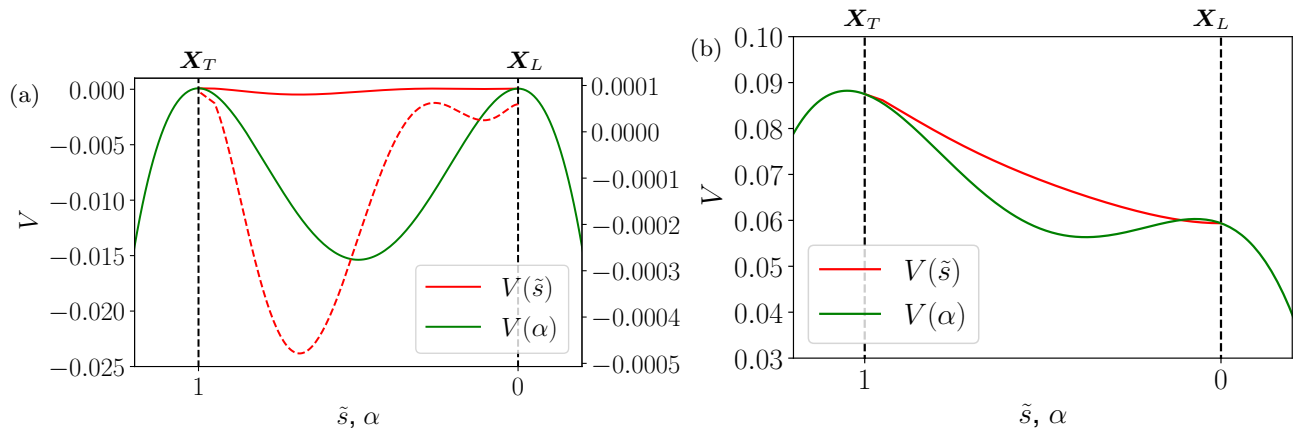
\begin{figure*}
    \centering
    \begin{overpic}[width=0.5\linewidth]{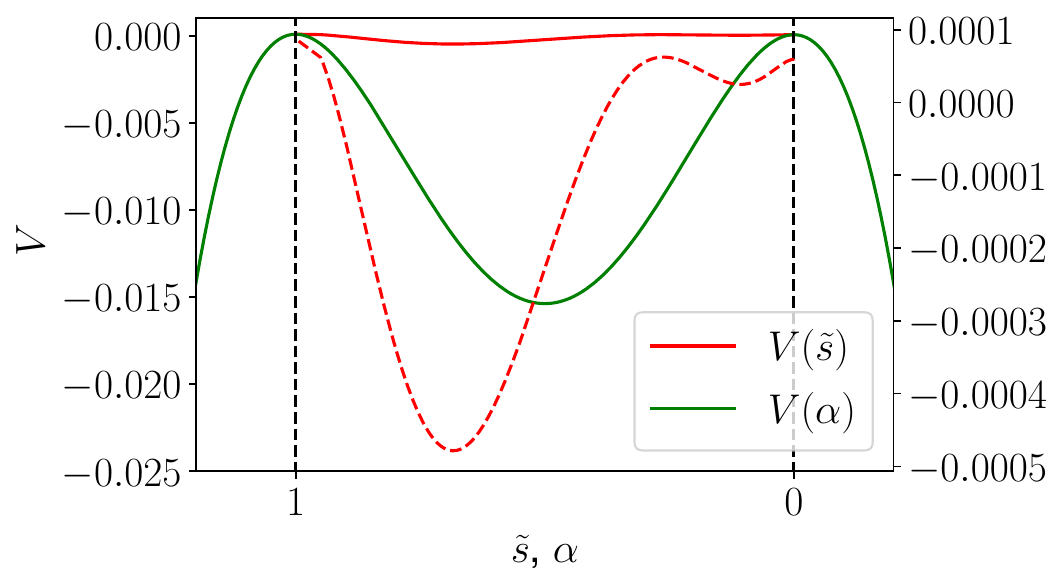}
    \put(0,50){(a)}
    \put(25,55){$\bm{X}_T$}
    \put(73,55){$\bm{X}_L$}
    \end{overpic}\begin{overpic}
    [width=0.45\linewidth]{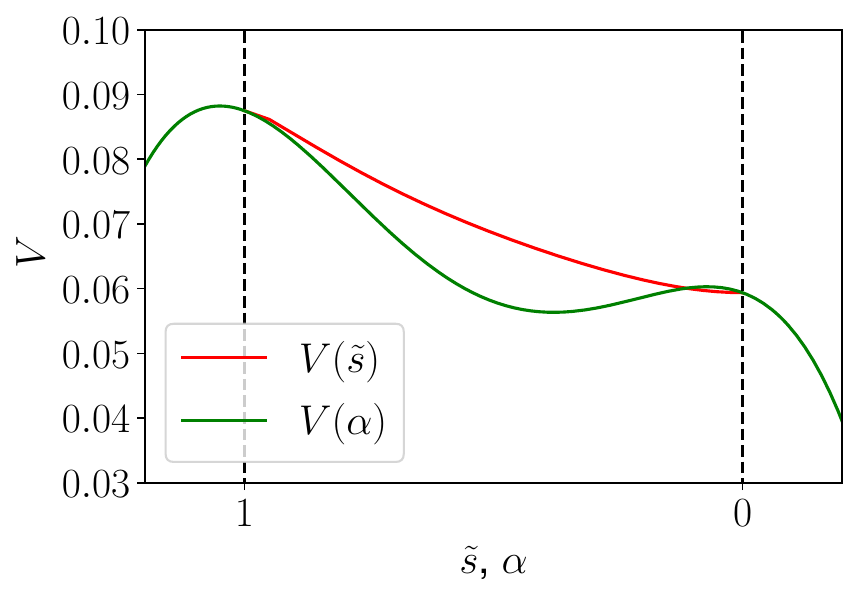}
    \put(0,63){(b)}
    \put(25,68){$\bm{X}_T$}
    \put(82,68){$\bm{X}_L$}
    \end{overpic}
    \caption{OM potential landscape $V$ [Eq.~\ref{eq:potential}] along the $\bm{Z}(\alpha)$ curve [red, $\tilde{s}\in[0,1]$] and the $\tilde{X}(\tilde{s})$ line [green, $\alpha\in\mathbb{R}$] for (a) $D=10^{-4}$ and (b) $D=10^{-1}$. The enlarged plot of the $\bm{Z}(\alpha)$ curve are shown in dashed red line in (a) (see the labels on the right perpendicular axis). The horizontal axis is flipped for better comparison with Fig.~\ref{fig:potential}.
    }\label{fig:V01}
\end{figure*}

We first numerically calculate the stationary PDF of the stochastic DM equation by Langevin simulations 
via the Euler-Maruyama scheme. The obtained PDFs $P(x,y)$ for different noise strengths, $D=\{10^{-4}, 10^{-3},10^{-2},10^{-1} \}$, are shown in Fig.~\ref{fig:pdf} with their maximum value normalized to unity and isolines of $P=0, 0.25, \dots, 1$ in color, where $3\times10^7$ samples were used for the plots.

When the noise strength is sufficiently small as in the $D=10^{-4}$ case [Fig.~\ref{fig:pdf}(a)], the distribution exhibits two peaks that correspond to the two stable fixed points of the DM model,
showing small fluctuations around the stable solutions. 
To further clarify the distribution structure, we introduce a line perpendicular to $\bm{Z}(\alpha)$ and label the perpendicular direction by $\beta$, leading to the marginal distribution $N(\alpha)$ which is defined as $N(\alpha)=\int P(\alpha,\beta)\,d\beta$. The  marginal distribution normalized by its maximum value, $\tilde{N}(\alpha)$, is shown in Fig.~\ref{fig:peaks}, which visualizes the PDF structure along the $\bm{Z}(\alpha)$ direction.

As $D$ increases to $D=10^{-3}$ [Fig.~\ref{fig:pdf}(b)], the peak around $\bm{X}_L$ disappears and the 
distribution shifts to a unimodal shape,
suggesting that $\bm{X}_L$ is no longer a stable state in the stochastic DM model. 
This qualitative change may be physically interpreted as follows: with a higher noise intensity, the tail of the distribution broadens to reach the separatrix, leaking into the opposite basin.
When we further increase $D$ to $D=10^{-2}$ [Fig.~\ref{fig:pdf}(c)], the distribution becomes even fatter and then covers the $\bm{X}_L$ fixed point, which is also visible in Fig.~\ref{fig:peaks}. 
At $D=10^{-1}$, a broad distribution then covers both $\bm{X}_L$ and $\bm{X}_T$ [Fig.~\ref{fig:pdf}(d)], and we found that the peak of the distribution is slightly away from $\bm{X}_T$ towards the $-y$ direction, 
which can also be seen in Fig.~\ref{fig:peaks}.

We then proceed to compare the PDF with the OM potential function $V$ [Eq.~\eqref{eq:potential}].  
Note that the noise strength can have an impact on MPPs only through the OM potential in Eq.~\eqref{eq:ELEM0}.
We plot the contour of the potential function $V(x,y)$ in Fig.~\ref{fig:potential} for $D=10^{-3}$, and $D=10^{-1}$, showing qualitative similarities between  $V(x,y)$  and $P(x,y)$ profiles, including positions of peaks and their dependence on the noise strength $D$.
To illustrate the positions of the potential extrema,
we further investigate the potential functions along the $\tilde{X}(\tilde{s})$ curve and the $\bm{Z}(\alpha)$ line, which are plotted in Fig.~\ref{fig:V01} for $D=10^{-3}$, and $D=10^{-1}$.

\begin{figure*}
    \centering
    \begin{overpic}[width=0.49\linewidth]{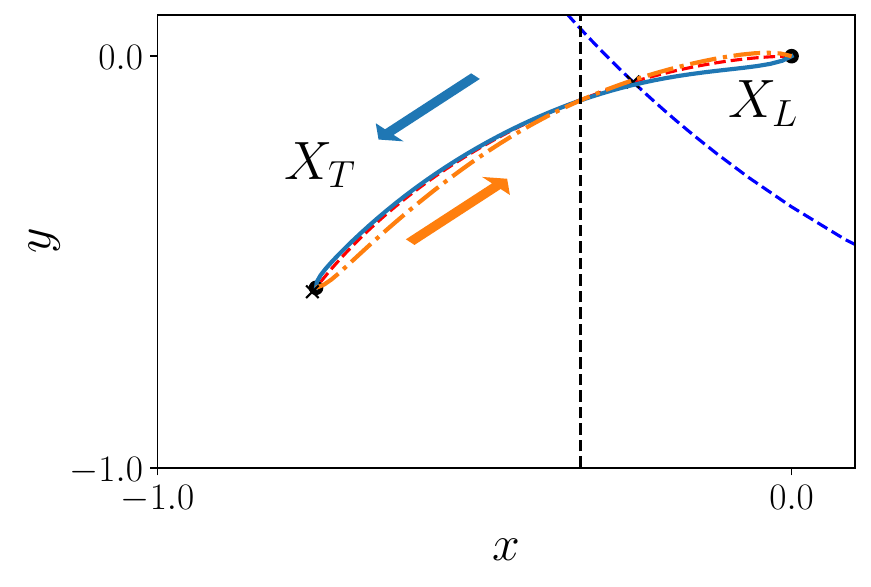}
    \put(4,63){(a)}
    \end{overpic}\begin{overpic}
    [width=0.49\linewidth]{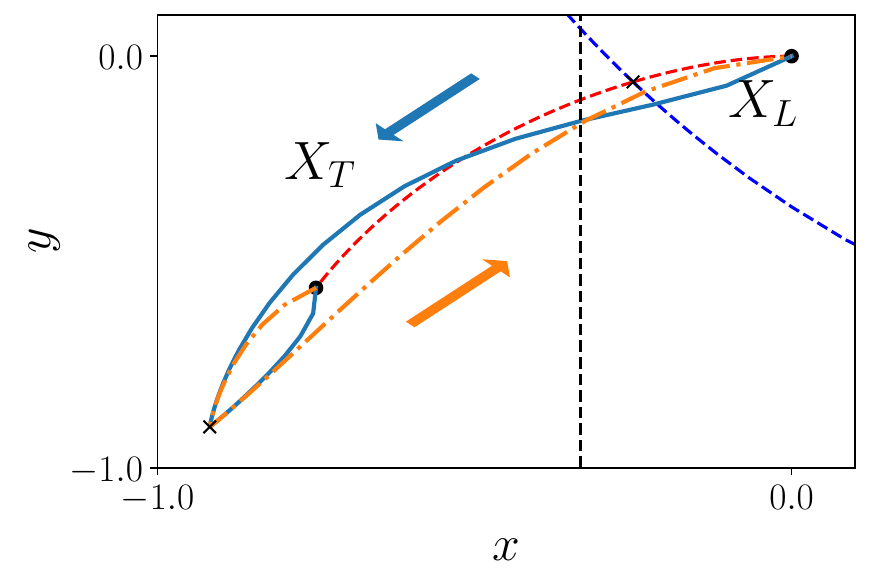}
    \put(4,63){(b)}
    \put(25,15){$\bm{X}_F$}
    \end{overpic}
    \caption{Most probable path connecting $\bm{X}_L$ and $\bm{X}_T$ for (a) $D=10^{-3}$ and (b) $D=10^{-1}$ with $\tau=30$.
    Solid (cyan) lines indicates turbulentization path and dotdash (orange) line indicates laminarization path.
    Vertical dashed line (black) shows $x=-1/3$, on which the sign of magnetic field reverses [Eq.~\eqref{eq:ELEM}]. The maximum point of the OM potential is denoted by $\bm{X}_F$ and marked by `$\times$' symbol.
    \label{fig:MPP}}
\end{figure*}
At $D=10^{-3}$, the $V(x,y)$ plots along the two curves are remarkably different in magnitude [Fig.~\ref{fig:V01}(a)] and 
the values along the $\tilde{\bm{X}}(\tilde{s})$ curve exhibit only minor variations relative to those observed along the $\bm{Z}(\alpha)$ line. 
This observation indicates that the curve $\tilde{\bm{X}}(\tilde{s})$, which is a heteroclinic orbit, 
behaves as a ridge in the OM potential function.
The enlarged plot of the potential function on the heteroclinic orbit is also shown in a dashed line in Fig.~\ref{fig:V01}(a), where one can find three local maxima. The middle one around $\tilde{s}\approx  0.265$ corresponds to the edge state $\bm{X}_E$, 
and the local maxima at $\tilde{s}=0$ and $\tilde{s}=1$ correspond to $\bm{X}_L$ and $\bm{X}_T$, respectively.

At $D=0.1$, the local maximum point along the $\bm{X}(\tilde{s})$ curve disappears, and its variation becomes comparable to 
that on the $\bm{Z}(\alpha)$ line. Interestingly, the maximum of the potential function is not on the fixed point $\bm{X}_T$, but slightly deviates from it [Figs.~\ref{fig:potential}(b),~\ref{fig:V01}(b)].
This qualitative shift of the OM potential function is remarkably comparable with the stationary PDF of the stochastic DM system. 

These qualitative agreements between the stationary PDF $P(x,y)$ and the potential landscape $V(x,y)$ are not just coincidences. As explained in details in Appendix~\ref{sec:schroedinger}, the time evolution of the  probability density function is described as a Euclidean Schr\"odinger equation under electromagnetic field whose scalar potential is $-V(x,y)$. Moreover, the stationary PDF $P(x,y)$ is obtained as the ground state of the Schr\"odinger equation, leading to a physical picture of the qualitative agreements between the landscapes of $P(x,y)$ and $V(x,y)$:
A peak of the potential landscape in Fig.~\ref{fig:potential} can then be read as potential confinements in the Schr\"odinger system.

In this section, we have seen that the OM potential function  inherits the invariant manifold of the original deterministic system as a ridge. Also, the stationary PDF of the stochastic dynamics is physically interpreted via the potential landscape with the analogy to the Schr\"odinger equation under electromagnetic field. In particular, the effects of the noise are well incorporated into the OM potential landscape. 
In the proceeding sections, we discuss the MPP and its relation with the OM potential function.

\section{Transition between the laminar and turbulent states and the most Probable paths\label{sec:mpp}}
\subsection{Most probable paths between laminar and turbulent states}
In this section, we investigate the MPP between the two stable fixed points, $\bm{X}_L$ and $\bm{X}_T$, physically representing the noise-induced transition between the laminar and turbulent states in the DM model. For latter use, we hereafter refer to the laminar-to-turbulent transition path,  $(\bm{X}_{\mathrm{init}},\bm{X}_{\mathrm{final}})=(\bm{X}_L,\bm{X}_T)$, as {\it turbulentization} path, 
and the turbulence-to-laminar transition path, $(\bm{X}_{\mathrm{init}},\bm{X}_{\mathrm{final}})=(\bm{X}_T,\bm{X}_L)$, as {\it laminarization} path.
As discussed in Sec.~\ref{sec:setting}, we only consider a single branch of MPP that is continued from the constant momentum solution at small $\tau$ and we have confirmed that other branches are not numerically found within the state space we examined.

At first, we show the transition path that connects $\bm{X}_L$ and $\bm{X}_{T}$  
at $D=10^{-3}$ with $\tau=30$ in Fig.~\ref{fig:MPP}.
In the case of $D=10^{-3}$,
the turbulentization and laminarization paths both follow (but are not exactly on) the heteroclinic orbit, where the ridge of the OM potential function $V$ is situated. The slight deviation of the MPP from the heteroclinic orbit reflects the rotational dynamics represented as magnetic field in Eq.~\eqref{eq:ELEM} and the two paths intersect at $x=-1/3$, where the strength of the magnetic field $|3x+1|$ vanishes.

\begin{figure*}
    \centering
    \begin{overpic}[width=0.32\linewidth]{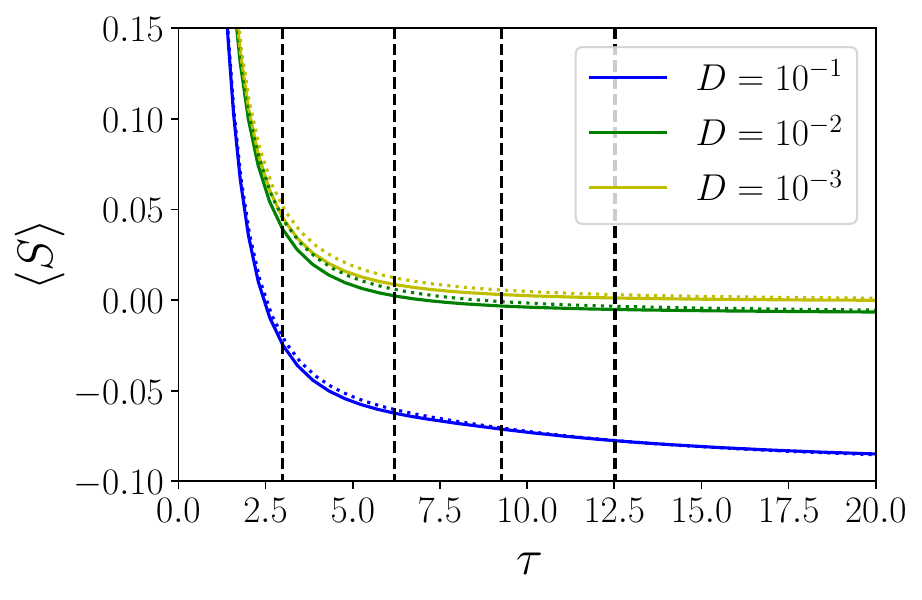}
    \put(1,58){(a)}
    \end{overpic}\begin{overpic}
    [width=0.32\linewidth]{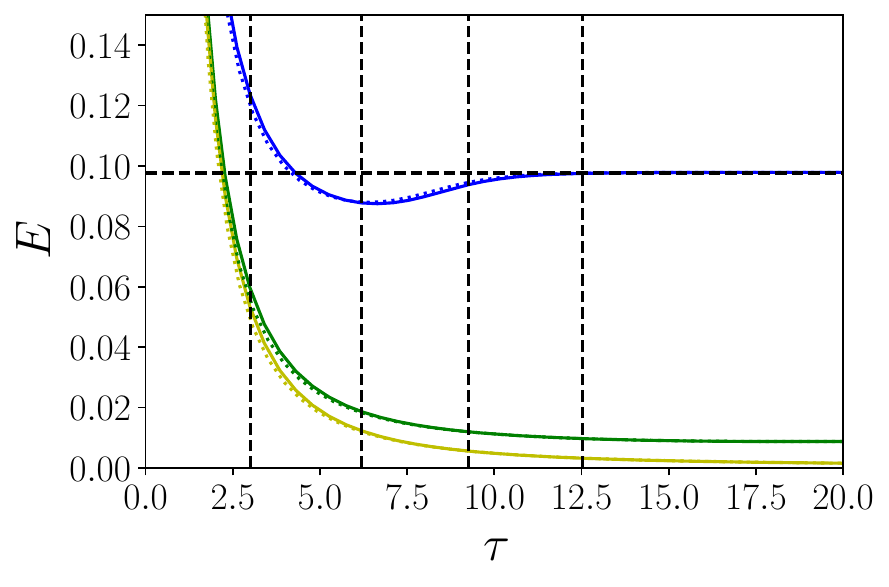}
    \put(0,58){(b)}
    \end{overpic}\begin{overpic}
    [width=0.32\linewidth]{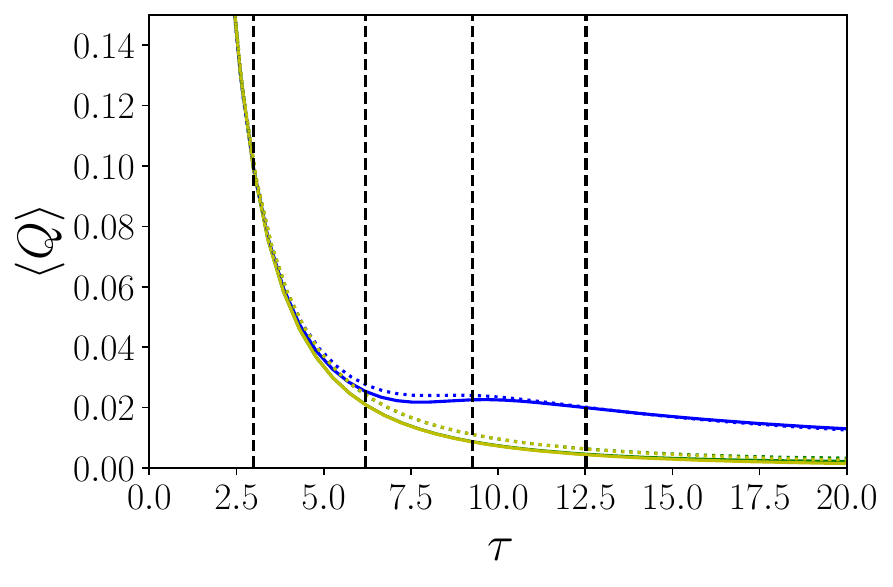}
    \put(0,58){(c)}
    \end{overpic}
    \caption{(a) Plots of action $\langle S \rangle=S/\tau$, (b) energy $E$, and (c) heat absorption $\langle Q\rangle=Q/\tau$ for $D=10^{-3}$, $10^{-2}$ and $10^{-1}$. 
    Solid and dotted lines indicate turbulization and laminarization transition, respectively.
    Four representative values of $\tau=\tau_i$ $(i=1,2,3,4)$, corresponding to the paths in Fig.~\ref{fig:typical_tau_path}, are shown in the vertical dashed lines. 
    }
    \label{fig:action}
\end{figure*}

\begin{figure*}
    \centering
    \begin{overpic}
    [width=0.49\linewidth]{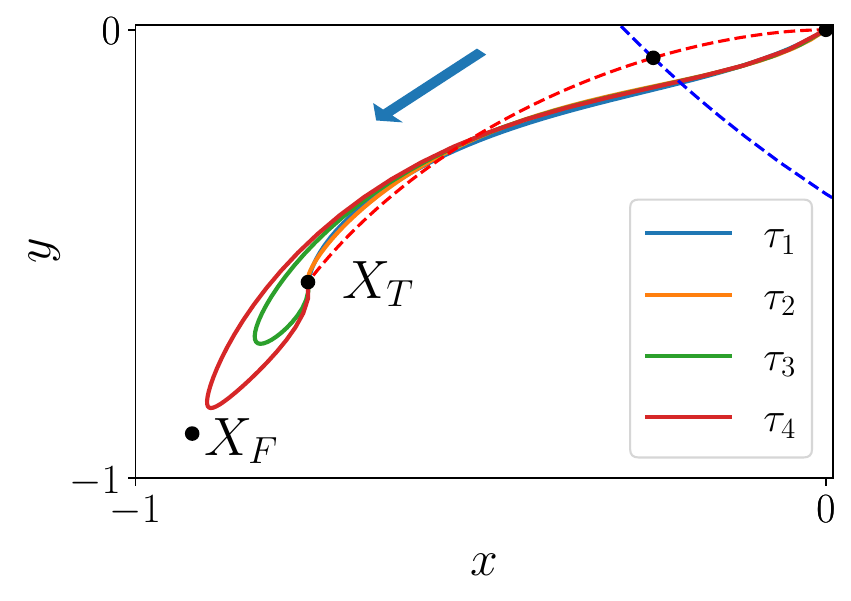}
    \put(3,64){(a)}
    \end{overpic}\begin{overpic}
    [width=0.49\linewidth]{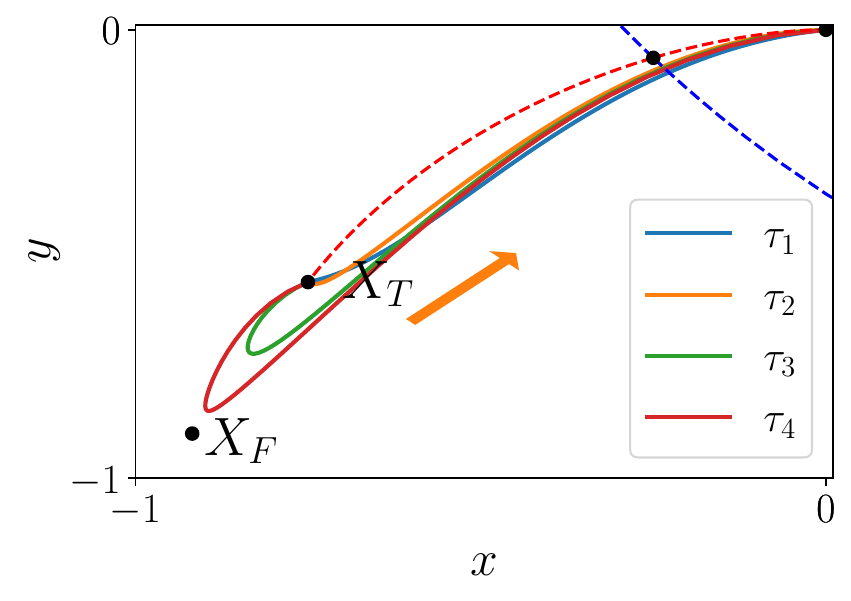}
    \put(3,64){(b)}
    \end{overpic}
    \caption{Transition paths of (a) turbulentization and (b) laminarization with different time duration $\tau=\tau_i$ $(i=1,2,3,4)$. 
    $\tau_1=3$ and $\tau_4=12.5$, and two other $\tau$s, $\tau_2$ and $\tau_3$, equally separate the time interval $[\tau_1, \tau_4]$.
\label{fig:typical_tau_path}}
\end{figure*}

When $D=10^{-1}$, the turbulentization and laminarization transition paths exhibit a remarkable nonreciprocity, as shown in Fig.~\ref{fig:MPP}(b). While the two transition paths significantly deviate from each other, both again intersect at $x=-1/3$.

As seen in the previous section, at a large $D$, the maximum of the OM potential function deviates from $\bm{X}_T$ [See also Fig.~\ref{fig:potential}(d)]. We denote the maximum of the OM potential function by $\bm{X}_F$ and mark as $\times$ symbol in Fig.~\ref{fig:MPP}(b). Both the turbulentization and laminarization transition paths transit at the $\bm{X}_F$ point before reaching the final state.

These results highlight the significant role of the OM potential landscape that characterizes the MPP and hence the transition pathway between the laminar and turbulent states, while the MPP follows the heteroclinic orbit of the original deterministic system as the ridge of the OM potential in a small $D$ case.

\subsection{$\tau$-dependence}

To obtain the MMP, one needs to fix the value of the transition time $\tau$. In a physical setup, however, the transition time is not a controllable parameter. 
In this subsection, we examine the $\tau$-dependence of the MPP between the laminar and turbulent states and discuss the qualitative changes in the MPP based on the kinematic formula of the MPP dynamics discussed in Sec.\ref{sec:mppkinematics}.

We have seen in Sec.~\ref{sec:mppkinematics} that the total energy is a conservative quantity from the Hamitonian structure of the MPP dynamics. We first calculate the energy of the MPP with different $\tau$-values as shown in Fig.~\ref{fig:action} for $D=10^{-3}$, $10^{-2}$, and $10^{-1}$.
We also show in the figure the OM action and heat absorption such that we have
\begin{align}
    \langle S\rangle=\langle Q\rangle-E
\end{align}
from Eq.~\eqref{eq:thermodynamics}, with $\langle S\rangle=S/\tau$ and $\langle Q\rangle=Q/\tau$.

\begin{figure*}
    \centering
    \begin{overpic}[width=0.40\linewidth]{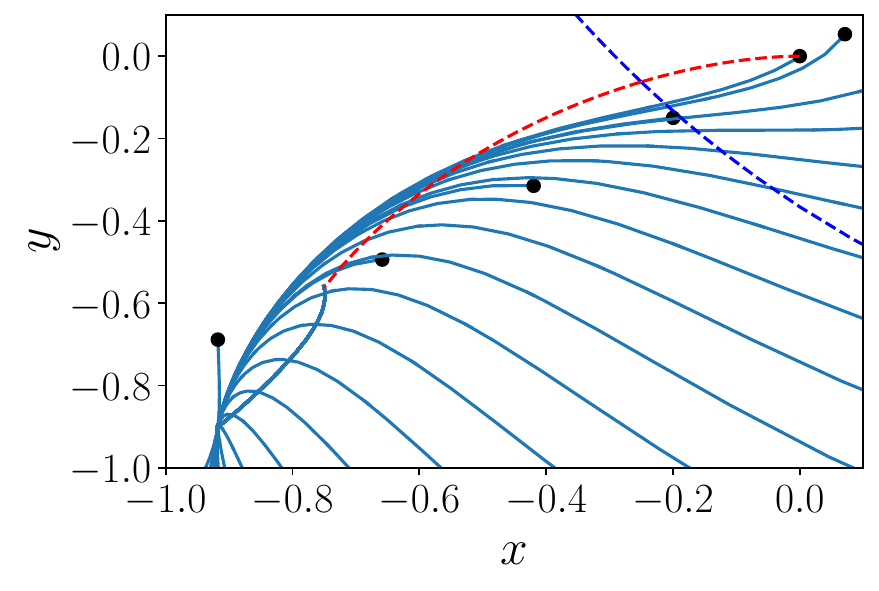}
    \put(3,62){(a)}
    \put(22,60){turbulentization}
    \end{overpic}\begin{overpic}
    [width=0.40\linewidth]{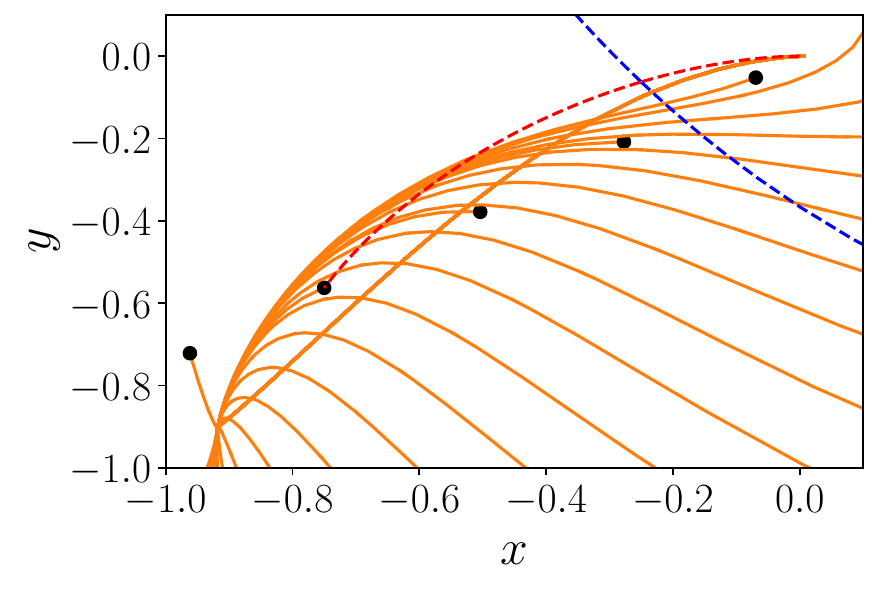}
    \put(3,62){(b)}
    \put(22,60){laminarization}
    \end{overpic} \\ \begin{overpic}
    [width=0.40\linewidth]{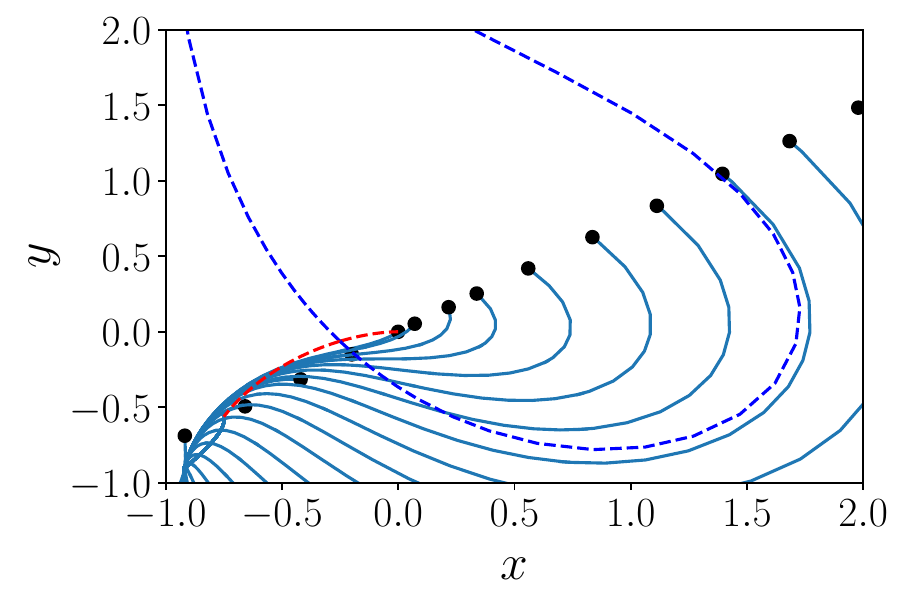}
    \end{overpic}\begin{overpic}
    [width=0.40\linewidth]{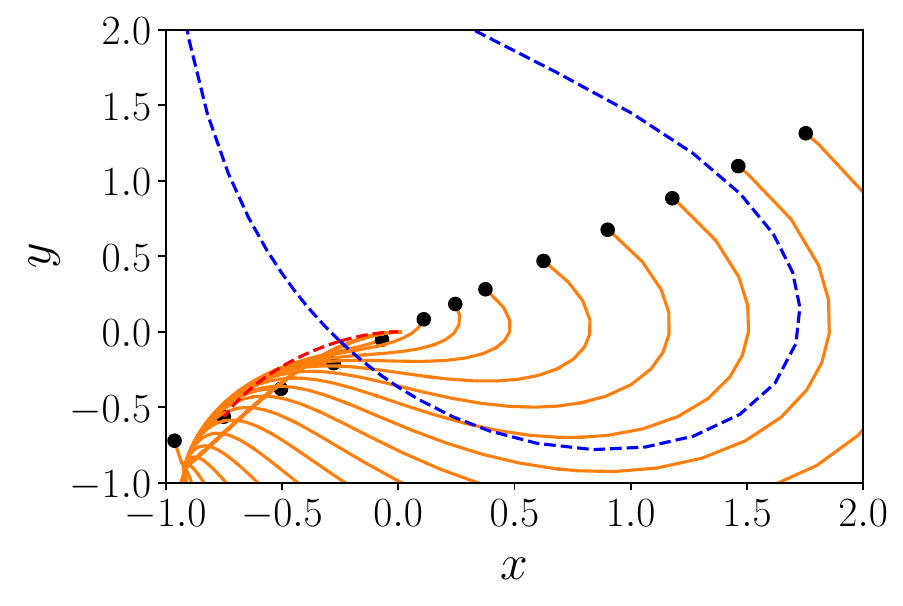}
    \end{overpic}
    \caption{Most probable paths with different initial states $X_{\rm init}$ for (a) turbulentization and (b) laminarization with $\tau=100$ and $D=0.1$. Initial states are marked by dots. Upper and lower panels show narrow and broad regions of the $(x,y)$-plane, respectively, as in Fig.~\ref{fig:phase}. 
    }
    \label{fig:posit_path}
\end{figure*}

As seen in Fig.~\ref{fig:action}(a), the action $\langle S\rangle =S/\tau$ is monotonically decreasing, regardless of the noise strength. 
Remarkably, in all the $\tau$ and $D$ cases, the action $S$ for the laminarization path exhibits a higher value than the corresponding the turbulentization path, 
indicating that turbulentization is more likely to occur than laminarization for a fixed transtion time [See Eq.~\eqref{relativeprop}].

Fig.~\ref{fig:action}(b) shows the energy $E(\tau)$ of the obtained MPP.
When $D=10^{-3}$ and $D=10^{-2}$, the energy $E$ also decreases monotonically as $\tau$ increases. when $D$ increases to $D=10^{-1}$, however, the energy $E(\tau)$ has a local minimum at $\tau\approx6$ and then asymptotically approaches a constant value $E\approx 0.1$ as $\tau$ further increases. This asymptotic value of $E$ is found to be identical to the maximum value of the potential function, $V(x,y)$. 
In all cases, no significant differences can be found between the turbulentization and laminarization paths.

The heat absorption $\langle Q\rangle=Q/\tau$ does not change at a small noise intensity ($D=10^{-3}$ and $D=10^{-2}$) and is monotonically decreasing  [Fig.~\ref{fig:action}(c)]. In the case of $D=10^{-1}$, however, the values of $\langle Q\rangle$ become significantly larger than those at a small $D$, and exhibit a non-monotonic behavior with a local minimum at $\tau\approx6$. Remarkably, the values are almost the same at a small $\tau\lesssim 5$, regardless of the noise strength.

These observations imply the existence of several characteristic timescales in the MPPs at $D=10^{-1}$. To further analyze these behaviors, we introduce four representative values of $\tau$, $\tau_i$ ($i=1,2,3,4$), as indicated by vertical dashed lines in Fig.~\ref{fig:action}. More precisely, we set $\tau_1=3$ and $\tau_4=12.5$ as a representative short and long $\tau$, with $\tau_2$ and $\tau_3$ equally separating the interval $[\tau_1, \tau_4]$. 

The corresponding MPPs for different $\tau_i$ are shown in Fig.~\ref{fig:typical_tau_path}.
In a short time scale ($\tau=\tau_1, \tau_2$), the transition path follows the heteroclinic orbit with a slight deviation due to the nonreciprocal magnetic field, both for the turbulentization [Fig.~\ref{fig:typical_tau_path}(a)] and laminarization [Fig.~\ref{fig:typical_tau_path}(b)] processes, similarly to Fig.~\ref{fig:MPP}(a). The paths in Fig.~\ref{fig:typical_tau_path} are almost overlapped for the $\tau=\tau_1$ and $\tau=\tau_2$ cases.

On long time scales, $\tau=\tau_3$ and $\tau_4$, where $E(\tau)\simeq V(\bm{X}_{F})$, 
the transition path temporarily resides around $\bm{X}_F$ before proceeding to the final state. 
The changes in the path shape are only observed at the vicinity of $\bm{X}_T$, and remarkably, we found that all the paths intersect with the separatrix almost at the same point, independent of the time scale $\tau$. Although the value of $\tau$ is not a controllable parameter in a physical system, the resulting transition paths possess some universal properties in common, which could be experimentally observable and will be further discussed in Sec.~\ref{sec:conclusion}.

\begin{figure*}
    \centering
    \begin{overpic}
    [width=0.32\linewidth]{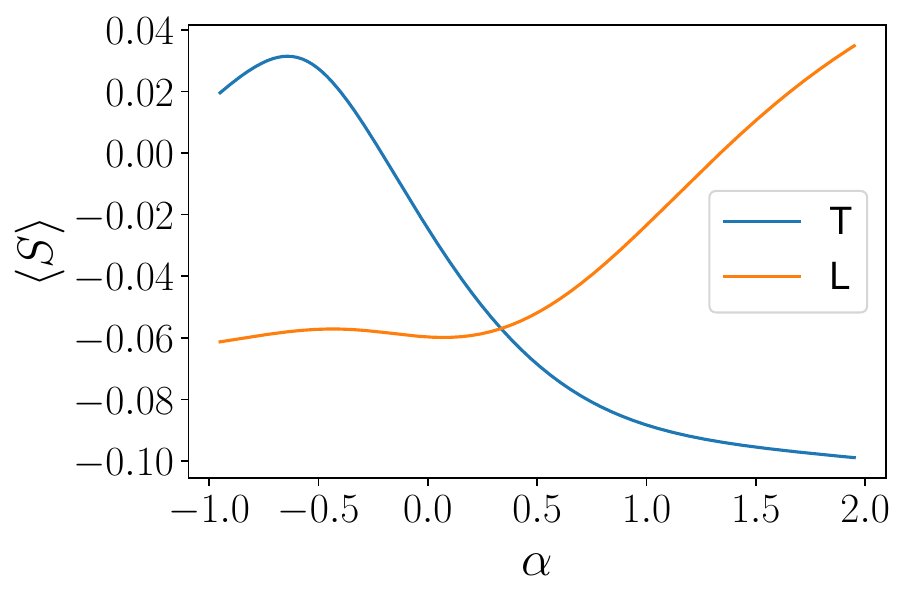}
    \put(0,62){(a)}
    \put(53,67){$\tau=\tau_1$}
    \end{overpic}
    \begin{overpic}
    [width=0.32\linewidth]{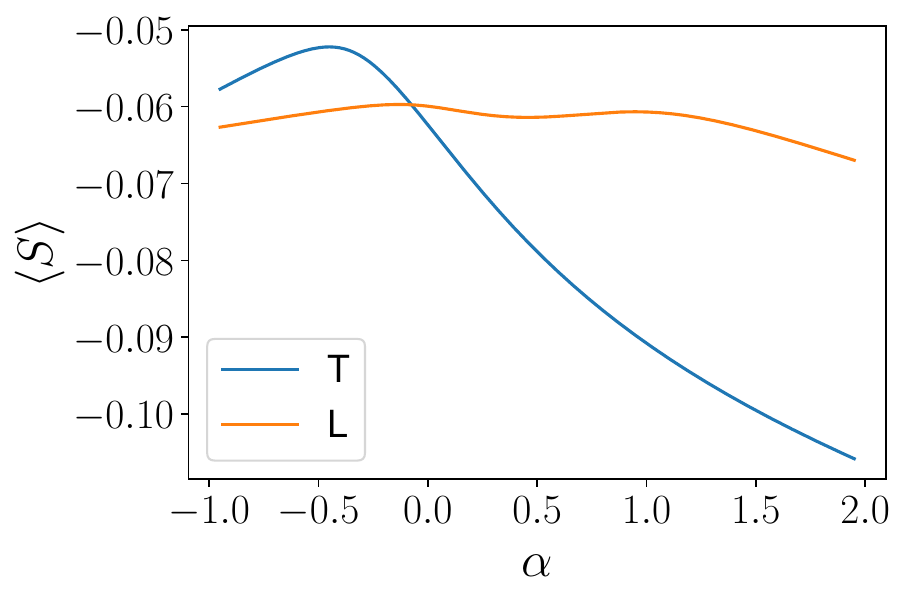}
    \put(0,62){(b)}
    \put(53,67){$\tau=\tau_2$}
    \end{overpic}
    \begin{overpic}
    [width=0.32\linewidth]{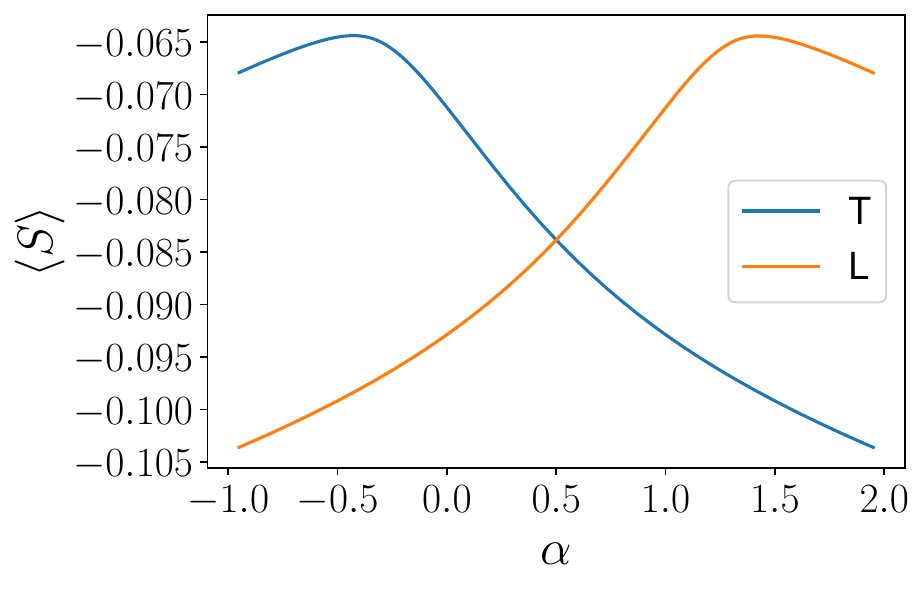}
    \put(0,62){(b)}
    \put(53,67){$\tau=\tau_3$}
    \end{overpic}
    \caption{ Plots of OM action $\langle S\rangle=S/\tau$ for different initial conditions $\bm{X}_{\textrm{init}}(\alpha)=\bm{Z}(\alpha)$ of the turbulentization (T, cyan) and laminarization (L, orange) paths with $D=10^{-1}$. The value of $\tau$ is set as (a) $\tau=\tau_1$, (b) $\tau=\tau_2$ and (c) $\tau=\tau_3$.
    }
    \label{fig:posit_prob}
\end{figure*}
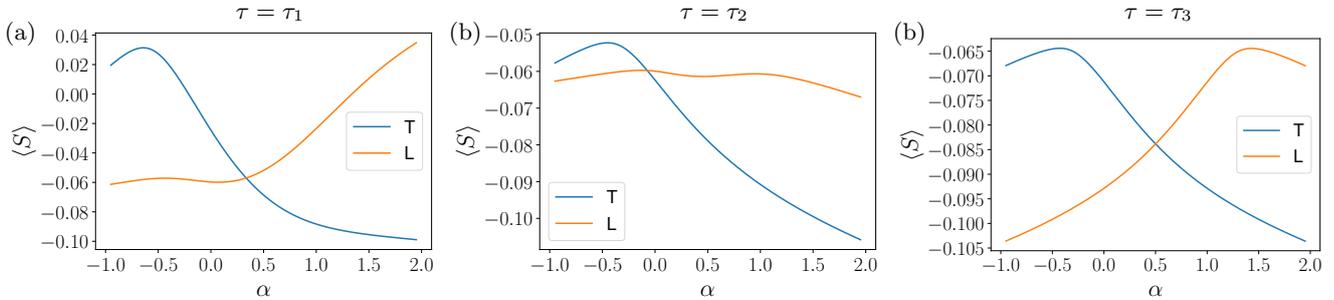

\subsection{Initial state dependence}

Finally, we examine the impact of the initial position on the MPP. 
Figure \ref{fig:posit_path} shows the obtained transition paths for the turbulentization and laminalization processes, where the initial states are selected from $\bm{X}_{\mathrm{init}}(\alpha)=\bm{Z}(\alpha)$ 
and we use $\tau= 100$. 

Since the time duration $\tau$ is relatively large $(\tau\gg \tau_4)$, all the obtained paths first approach $\bm{X}_F$ and reside around it both for the turbulentization and laminalization processes. The circulatory MPP trajectories mainly follow the deterministic flow of the DM model until they reach the neighborhood of $\bm{X}_F$. 
After leaving the neighborhood of $\bm{X}_F$, all the obtained paths trace almost the same trajectories to reach the final state $\bm{X}_{\mathrm{final}}$, highlighting that the initial-state dependence is only visible until the trajectories reach the $\bm{X}_F$ point.

Next, we evaluate the value of the OM action, $S$ with different initial states $\bm{X}_{\mathrm{init}}(\alpha)=\bm{Z}(\alpha)$. In Fig.~\ref{fig:posit_prob}(a), plots are shown against different $\alpha$ values both for the turbulentization path $(\bm{X}_{\rm final}=\bm{X}_T)$ and for the laminarization path $(\bm{X}_{\rm final}=\bm{X}_L)$. Each panel shows the results 
with $\tau=\tau_1$ [FIG.\ref{fig:posit_prob}(a)], $\tau=\tau_2$ [FIG.\ref{fig:posit_prob}(b)], and $\tau=\tau_3$ [FIG.\ref{fig:posit_prob}(c)]. 

In each $\tau$, we found that the action $S$ has lower values when starting near the $\bm{X}_T$ point ($\alpha=1)$, indicating that the turbulentization is more likely if the initial state is close to $\bm{X}_T$. 
This intuitive process, however, does not occur for the laminalization path, 
as seen from the flat plot of $S$ at $\tau=\tau_2$.

We then introduce the difference of the action between the turbulentization, $S_T$, and laminarizaton, $S_L$, as $\Delta S =S_T-S_L$, which quantifies the relative path probability between the two transitions with a fixed initial state [Eq.~\eqref{eq:dS}]. 
A positive value of $\Delta S$ indicates that the probability of transition to turbulence is higher than that of laminarization. The critical value of $\alpha$ on which the two transition probabilities become equal can be found as the intersection of the plots of actions in Fig.~\ref{fig:posit_prob}. 
The critical $\alpha$ value moves from $\alpha^\ast \approx 0.335$ at $\tau=\tau_1$ to  $\alpha^\ast\approx -0.0743$ at $\tau=\tau_2$ and $\alpha^\ast\approx 0.506$ at $\tau=\tau_3$, suggesting the existence of multiple timescales in the transition process. From these observations, we find that the turbulentization transition is enhanced at the intermediate timescale around $\tau=\tau_2$.

\section{Bifurcation of fixed points of Euler-Lagrange equation\label{sec:dynsys}}
In the previous section, we have seen that the stationary solution of the MPP dynamics, denoted by $\bm{X}_F$, played a significant role in determining the qualitative behavior of the MPP. In this section, to further characterize the MPP, we  examine steady state solutions for the MPP dynamics and their bifurcation diagram with different noise strength. 

\begin{figure*}
    \centering
    \begin{overpic}
    [width=0.49\linewidth]{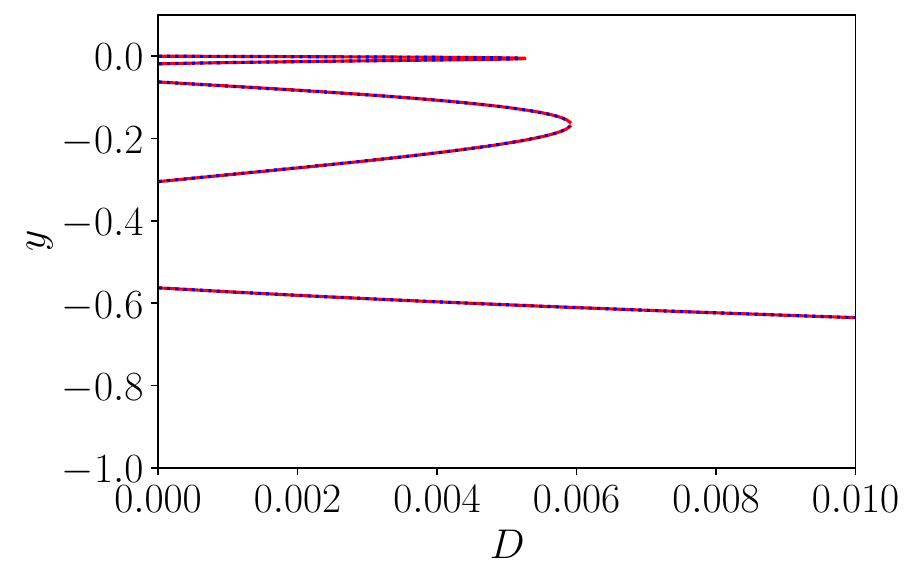}
    \put(1,58){(a)}
    \put(40,59){$\swarrow$}
    \put(43,62){$\bm{W}_L$}
    \put(35,54){$\nearrow$}
    \put(30,50){$\overline{\bm{W}}_L$}
    \put(60,53){$\swarrow$}
    \put(64,54){$\bm{W}_E$}
    \put(58,45){$\nwarrow$}
    \put(62,43){$\overline{\bm{W}}_E$}
    \put(70,31){$\swarrow$}
    \put(74,33){$\bm{W}_T$}
    \end{overpic}\begin{overpic}
    [width=0.49\linewidth]{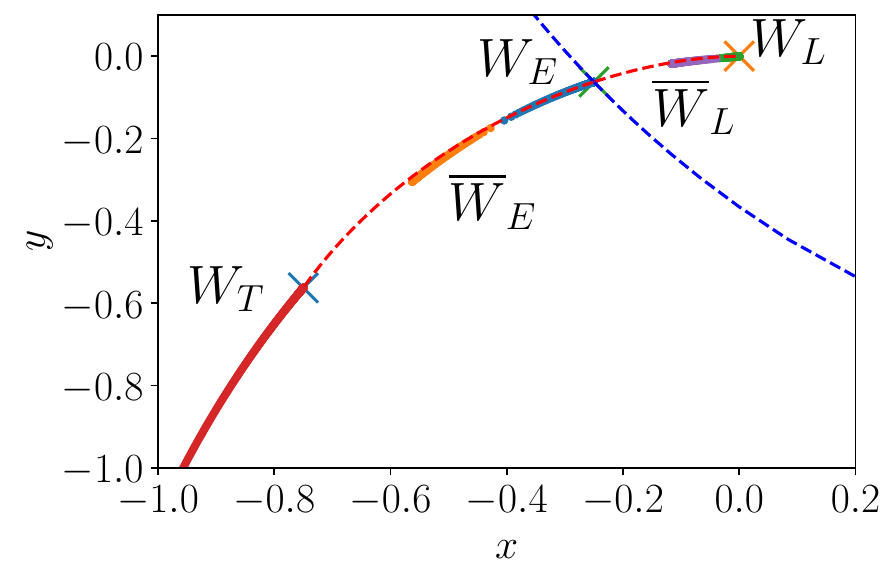}
    \put(1,58){(b)}
    \end{overpic}
    \caption{(a) Bifurcation curve of fixed points of the Euler--Lagrange equations in the DM model as a function of $D$. 
    Solid and dotted lines indicate fixed points and stationary points, respectively, which perfectly agrees as expected.
    (b) Position of the five fixed points superpositioned on the invariant set of the deterministic DM system for $D\in[0,0.1]$. 
    }
    \label{fig:bifurcation}
\end{figure*}
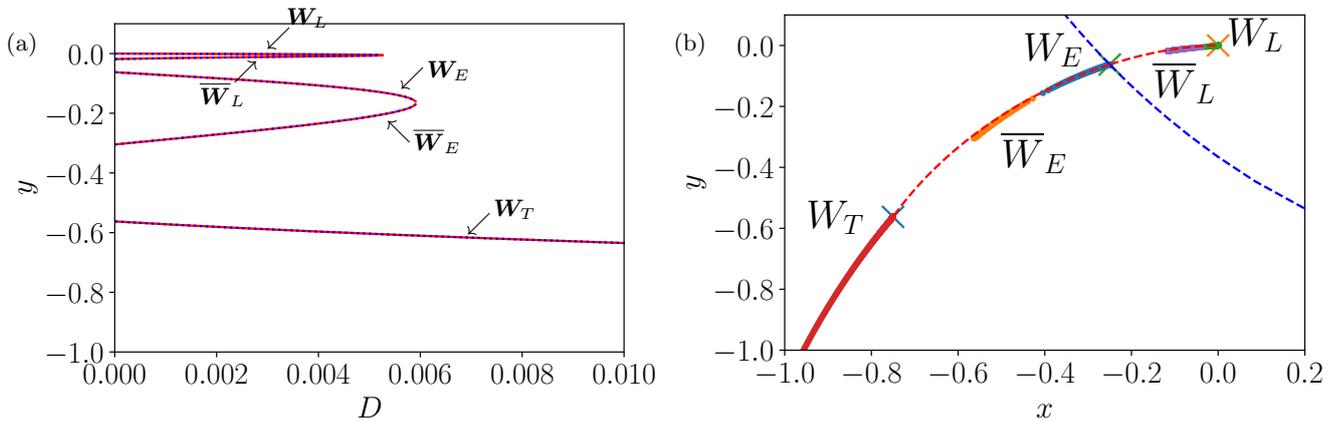

Figure~\ref{fig:bifurcation}(a) shows obtained bifurcation curves of fixed points for the EL equations [Eqs.~\eqref{eq:EL1}-\eqref{eq:EL2}]. 
At $D=0$, we found five fixed points corresponding to the extrema of the OM potential [See also Fig.~\ref{fig:V01}(a)]. 
Three of the fixed points are identical to the fixed points of the deterministic DM model and the other two fixed points have a finite momentum $(\bm{x}, \bm{p})=(\bm{x}^\ast,-\bm{f}(\bm{x}^\ast))$. 
We refer to the solutions in the branch connecting to $\bm{X}_L$, $\bm{X}_E$, and $\bm{X}_T$ as 
$\bm{W}_L$, $\bm{W}_E$, and $\bm{W}_T$, respectively.
Moreover, two finite-momentum solutions are denoted by $\overline{W}_L$ and $\overline{W}_E$

As seen in Fig.~\ref{fig:bifurcation}(a), fixed points $\bm{W}_L$ and $\bm{W}_{E}$ disappear at $D\approx 0.005$ by 
colliding with $\overline{\bm{W}}_L$ and $\overline{\bm{W}}_E$, respectively. 
We also numerically found that the fixed point $\bm{W}_{T}$ survives at an arbitrarily large $D$.
Moreover, the fixed point $\bm{W}_{T}$ is found to be identical to 
$\bm{X}_F$, the maximum point of the OM potential function. 
The OM potential landscape significantly changes around the bifurcation point, $D\approx0.005$.

Here, we analyze the fixed points of the Euler–Lagrange equation to understand the shape of the transition path.
Figure \ref{fig:bifurcation}(b) shows the location of the fixed points, together with the invariant set of the deterministic DM model. 
Notably, four fixed points $\bm{W}_L$, $\bm{W}_E$, $\overline{\bm{W}}_L$, and $\overline{\bm{W}}_E$ are found to be located in the invariant set, even with a nonzero $D$. The other fixed point, $\bm{W}_{T}$, deviates from $\bm{X}_{T}$ towards $-y$ direction as $D$ increases, being consistent with the OM potential landscape in Fig.~\ref{fig:potential}(b).

From the bifurcation diagram, we come to a clear picture that at a large $D$ ($D\gtrsim 0.005$) after the final bifurcation the MPP trajectory consists of three parts: (1) a relatively straight path from $\bm{X}_{\rm init}$ to $\bm{X}_F$, (2) residence at $\bm{X}_F$, and (3) a relatively straight path from $\bm{X}_{F}$ to $\bm{X}_{\rm final}$.

Here, the morphology of the MPP is governed only by the landscape of the OM potential function $V(\bm{x})$, which is straightforwardly obtained from the original deterministic system by Eq.~\eqref{eq:potential0}, even for a general system with a large degrees of freedom. Recalling that solving the MPPs requires careful treatments in computing the boundary-value problem even in the two-dimensional model system as explained in Sec.~\ref{sec:numeric}, the OM potential function provides a systematic pathway to understand the stochastic system with invariant sets of the original deterministic problem.

\section{concluding remarks\label{sec:conclusion}}
In this study, we investigated the most probable path 
of noise-induced transitions in a two-variable model of subcritical transition.
We formulated the Euler-Lagrange equations from minimizing the Onsager–Machlup action and found that the most probable path follows the Hamiltonian equation under electromagnetic field with an effective potential function, which we introduced as the Onsager-Machlup potential. We found that the OM potential is useful in qualitatively estimating the stationary probability density of the stochastic dynamics and also useful in understanding the transition path between laminar and turbulent states.

We numerically demonstrated that the most probable paths capture the nonreciprocal transition probabilities between the laminar and turbulent states. The morphology of the most probable path with a fixed duration time $\tau$ is analyzed through a kinematic relation with energy and heat absorption, which is derived from the Hamiltonian structure.
By evaluating the energy function, we have found that the transition dynamics varies depending on the transition time  $\tau$, along with the deviation of the OM potential maxima from the turbulent fixed point. We also found that the obtained OM action for the turbulentiation is always lower than the laminarization process regardless of the value of $\tau$, indicating that the turbulentization transition is found to be more frequent than the laminarization transition in the DM model. 
 
The Hamiltonian structure also enables us to 
rewrite the solutions of the EL equation of the model as the extrema of the OM potential function of the noise-induced system. Remarkably, we have found that 
these extrema inherit the invariant set of the deterministic system, and that the noise-induced transition paths therefore follow the OM potential landscape and its bifurcation diagram with the noise strength.

In high-dimensional systems such as fluid flows, noise-induced dynamics is generally nonreproducible, and obtaining the probability density function, a fundamental quantity of stochastic motion, is almost hopeless.
In contrast, methods for computing invariant solutions in dynamical systems have been extensively developed in fluid mechanics and have been successful in capturing the essential motions of the Navier–Stokes equations \cite{kawahara_2012}. 
The present results based on the Hamiltonian structure of the OM action are straightforwardly applicable to such a high-dimensional system and could provide a theoretical basis to predict noise-driven dynamics from deterministic systems.

In this study, fixed points of the EL equation, or equivalently the extrema of the OM potential, have never been found on the separatrix or the stable manifolds 
of the edge state, in contrast to the turbulent transition scenario proposed by infinitesimally small noise \cite{Wan_2015}. 
Rather, by using the OM formulation for a finite noise strength $D$, we numerically demonstrate that all the transition paths intersect with the separatrix at the same point away from the edge state, supporting the transition scenario at a finite noise \cite{rolland2024}.
Moreover, we numerically confirmed that the transition paths converge regardless of transition time $\tau$ and initial state. Hence, the transition scenario could be experimentally and numerically reproducible in physical systems including fluid turbulence, although the noise-induced transition is, in general, not controllable.

The noise-induced turbulentization and re-laminarization processes have contributed to the understandings of turbulence transition as a nonequilibrium phase transition, 
and examples may include recent developments on directed percolation universality class \cite{sano2016universal,lemoult2016directed,hiruta_2020,tuckerman2020,hiruta2022,gome2022}. 
In a lattice model of directed percolation, such phase transition is triggered by stochastic and local interactions between active and inactive sites. In contrast, the governing dynamics of fluid flow is essentially deterministic and contains spatially nonlocal interactions through the pressure field.
Our theoretical approach, consistently covering invariant sets of a deterministic system and its stochastic behavior under a finite-size noise, could bridge the two different aspects of turbulence transition.

Bistability is not limited to turbulent transitions and  various other fluid systems have been known to exhibit multiple coexisting stable states, 
such as the multiple states in Taylor-Couette flow \cite{Huisman2014}, flow reversals of large-scale flows \cite{sugiyama2010,Mishra2015,hiruta_2017,suri2024}, and localized structure in bioconvection \cite{shoji2014,yamashita2025,hiruta2025bc}. 
Apart from the Navier-Stokes flow, our formulation based on the OM action could be potentially useful in mesoscopic phenomena including active fluids
\cite{dunkel2013,hiruta2024,nishiguchi2025}, where model equation is not usually derived from the first principles. 

While the current subcritical transition model is restricted to two dimensions, 
dynamics in a high-dimensional phase space, including chaotic behavior, should be significant 
in a more realistic and faithful representation of turbulent systems. As presented in this study, the OM potential is accessible without computational difficulties even for a high-dimensional system.
Future work therefore aims to focus on such large-dimensional nonlinear systems,
and we expect our approach to work even for developed turbulence.

\begin{acknowledgments}
Y.H., K.Y. and K.I. acknowledge the Japan Society for the Promotion of Science (JSPS) KAKENHI (Grant No. 21H05309);
K.Y. acknowledges JSPS KAKENHI (Grant No. 25K17357);
K.I. acknowledges JSPS KAKENHI (Grant No. 24K21517) and the Japan Science and Technology Agency (JST), FOREST (Grant No. JPMJFR212N) and CREST (Grant No. JPMJCR25Q1). 
\end{acknowledgments}

\appendix
 \section{Fokker-Planck equation as a non-Hermitian Sch\"odinger system}\label{sec:schroedinger}

The Fokker-Planck equation for the Langevin dynamics in Eq.~\eqref{eq:langevin} provides the time evolution of the probability density function $P(\bm{x}, t)$ in the form,
\begin{align}
    \frac{\partial P}{\partial t}=-\nabla\cdot\bm{J},
\end{align}
with probability current field,
\begin{align}
    \bm{J}(\bm{x})=P(\bm{x})\bm{f}(\bm{x})-\frac{D}{2}\nabla P(\bm{x}).
\end{align}
Here, the Fokker-Planck Hamiltonian, defined as $\partial_tP=-\mathcal{H}_{\rm FP}/D$, is written as 
\begin{align}
    \mathcal{H}_{\rm FP}=-\frac{D^2}{2}\left[\nabla-\bm{A}(\bm{x})\right]^2-V(\bm{x}),
    \label{HamiltonianFP}
\end{align}
where the vector field $\bm{A}(\bm{x})$ is introduced as $\bm{A}(\bm{x})=\bm{f}(\bm{x})/D$
and the scalar potential $V(\bm{x})$ exactly coincides with the OM potential function introduced in  Eq.~\eqref{eq:potential0}. 
In the context of Fokker-Planck equation, this potential function is called as an effective potential \cite{risken1989}, although a gradient field is usually assumed as $\bm{f}(\bm{x})=-\nabla U(\bm{x})$ in the Fokker-Planck context. 

The Fokker-Planck Hamiltonian [Eq.~\eqref{HamiltonianFP}] is interpreted as Euclidean non-Hermitian quantum Hamiltonian under electromagnetic fields \cite{mazzolo2023nonequilibrium}. The corresponding Scr\"odinger equation is read as a particle motion in an electromagnetic field with the particle mass $m=1$, the Planck constant $\hbar=1/D$, the scalar potential $V(\bm{x})$ and a purely imaginary vector potential $-i\bm{A}(\bm{x})$.

Suppose that the probability current $\bm{J}$ vanishes at exterior boundaries (including infinite far-field), stationary probability density function $P(\bm{x})$ exists and is formulated as
\begin{align}
    \mathcal{H}_{\rm FP}P=0.
\end{align}
Hence, the stationary PDF is obtained as the right-eigenfunction of the zero-energy mode, which provides the ground state. Note here that the eigenfunctions are not orthogonal due to the non-Hermitian Hamiltonian.

The current field then holds the divergence-free condition, $\nabla\cdot\bm{J}=0$ at the steady state, while the current does not necessarily vanish. To further look into this effect, let us decompose the dynamics into reversible and irreversible parts as $\bm{f}=\bm{f}_{\rm rev}+\bm{f}_{\rm irr}$, where $\bm{f}_{\rm rev}$ is written as a gradient of a potential $\bm{f}_{\rm rev}=-\nabla U$ and $\bm{f}_{\rm irr}=\bm{f}-\bm{f}_{\rm rev}$. When the detailed balance is satisfied, $\bm{f}_{\rm irr}=\bm{0}$ holds and the stationary PDF then follows $P(\bm{x})\propto \exp(-U(\bm{x})/D)$ and $\bm{J}=\bm{0}$. When the detailed balance is broken, however, the probability current remains as $\bm{J}(\bm{x})=P(\bm{x})\bm{f}_{\rm irr}(\bm{x})$, leading to nonreciprocal transition paths \cite{yasuda2022time,ishimoto2023odd}. The magnetic field associated with the vector potential $\bm{A}$ also remains nonzero as $\nabla\times\bm{A}=\nabla\times\bm{f}_{\rm irr}/D$, which should vanish in the reversible case.

Despite the nonzero current field, the landscape of the stationary PDF is essentially obtained as the ground state of the Sch\"oredinger equation under the scalar potential $V(\bm{x})$. Hence, the peaks of the potential landscape in Fig.~\ref{fig:potential} can be read as potential confinements in the Fokker-Planck Hamiltonian, yielding  peaks of stationary PDF $P(\bm{x})$.

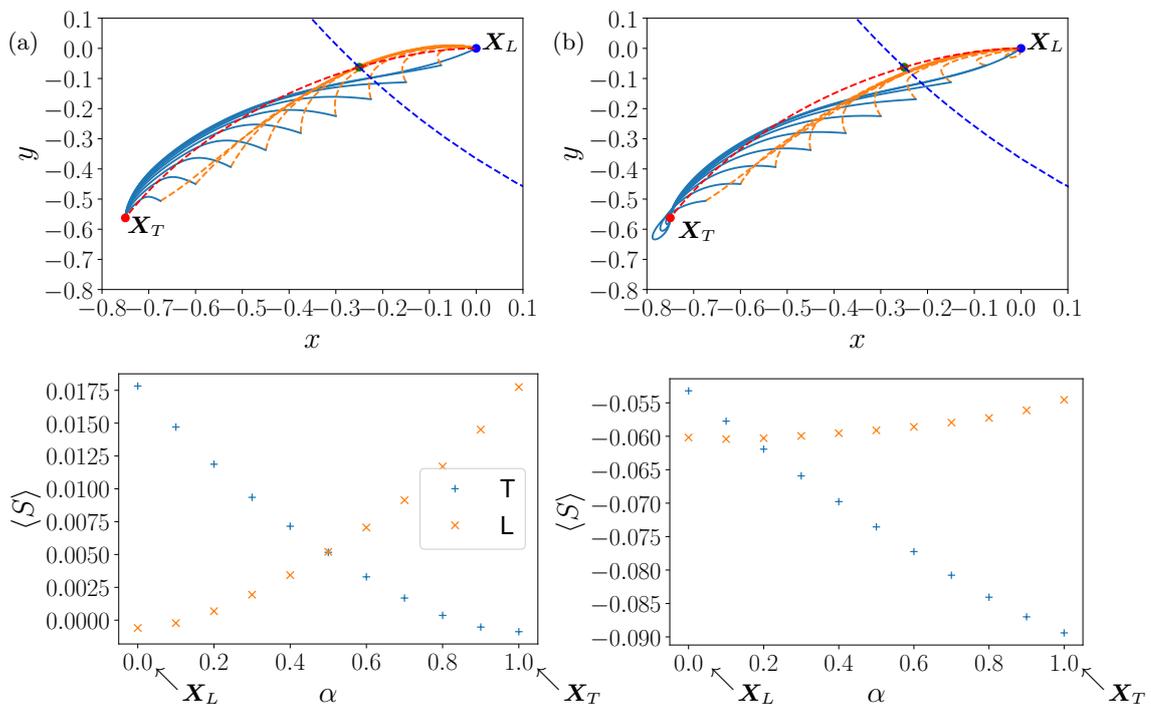
\begin{figure*}
    \centering
    \begin{overpic}
    [width=0.4\linewidth]{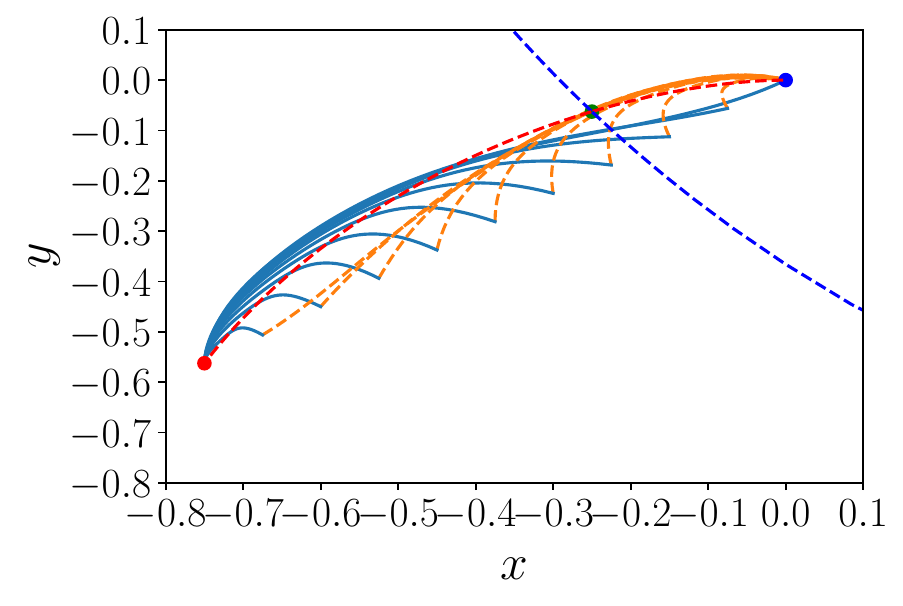}
    \put(1,58){(a)}
    \put(23,24){$\bm{X}_T$}
    \put(88,58){$\bm{X}_L$}
    \end{overpic}\begin{overpic}
    [width=0.4\linewidth]{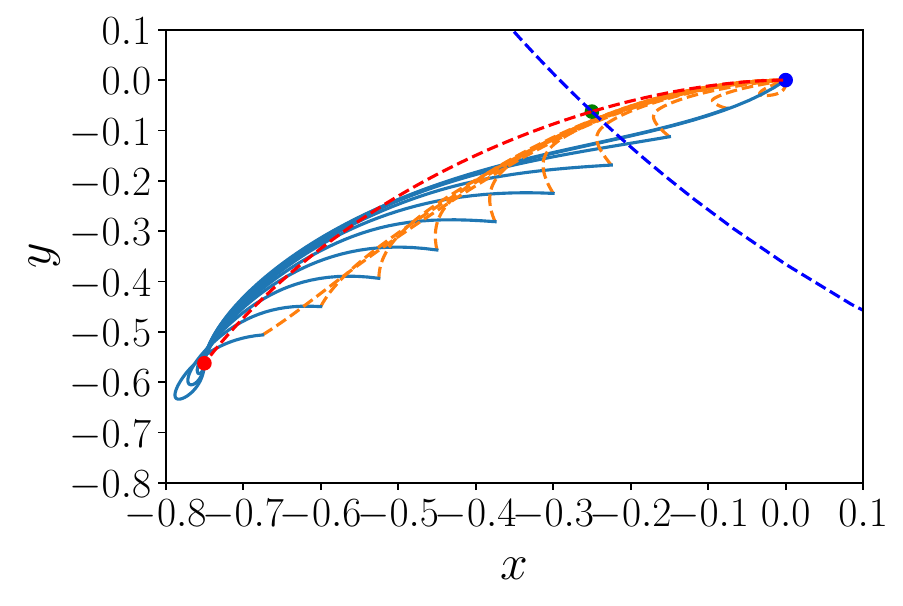}
    \put(1,58){(b)}
    \put(24,23){$\bm{X}_T$}
    \put(88,58){$\bm{X}_L$}
    \end{overpic}\\
    \begin{overpic}[width=0.4\linewidth]{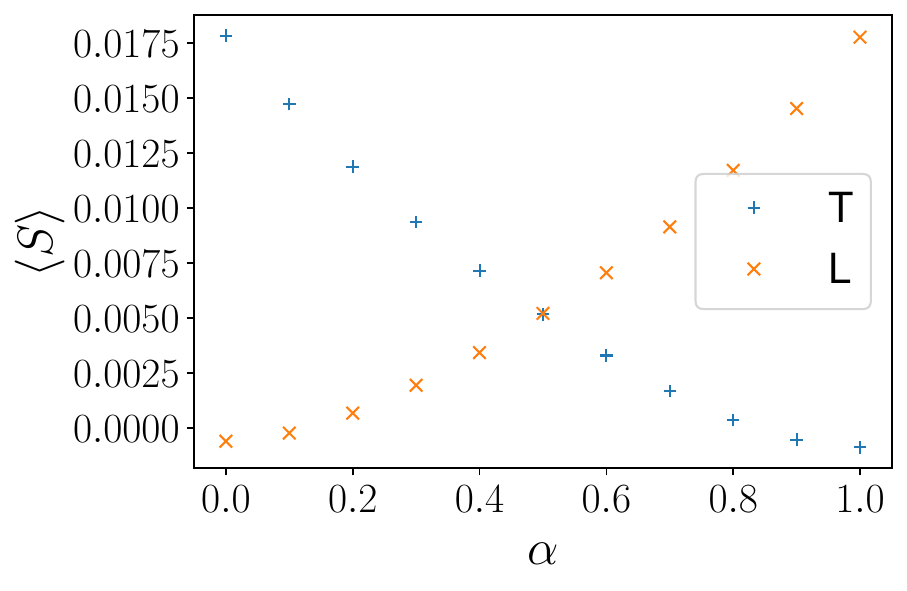}
    \put(28,6){$\nwarrow$}
    \put(98,6){$\nwarrow$}
    \put(33,3){$\bm{X}_L$}
    \put(103,3){$\bm{X}_T$}
    \end{overpic}\begin{overpic}
    [width=0.4\linewidth]{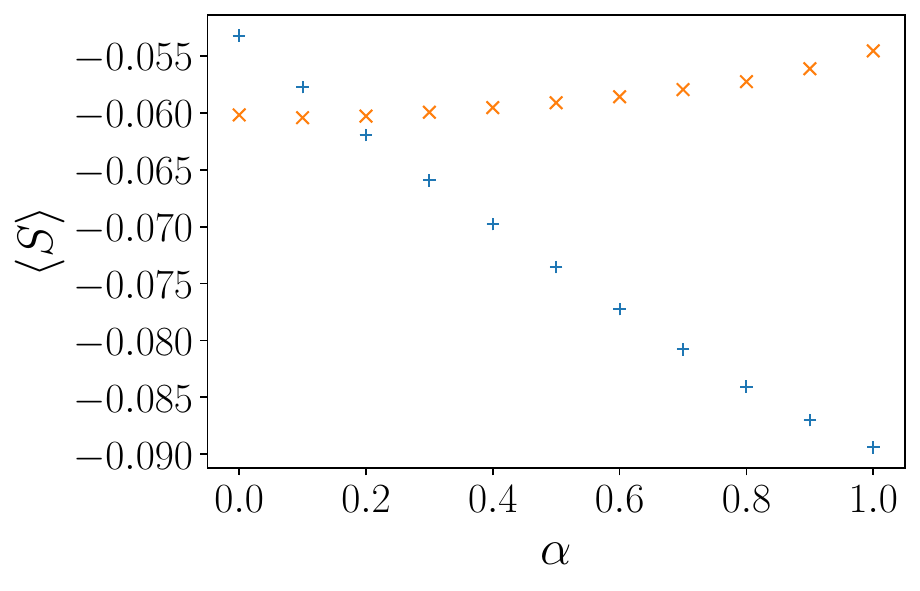}
    \put(30,6){$\nwarrow$}
    \put(98,6){$\nwarrow$}
    \put(35,3){$\bm{X}_L$}
    \put(103,3){$\bm{X}_T$}
    \end{overpic}
    \caption{MPPs (top) and the associated action value (bottom) with different initial states $\bm{X}_{\textrm{init}}=\bm{Z}(\alpha)$ 
    for (a) $D=10^{-3}$ and (b) $D=10^{-1}$ with $\tau =5$. As in Figs.~\ref{fig:MPP}, \ref{fig:posit_path} and \ref{fig:posit_prob}, turbulentization (T, cyan) and laminarization (L, orange) transitions are shown.
    }
    \label{fig:posit_probA}
\end{figure*}

 \section{Searching for the most probable path using deep neural network\label{sec:neural}}
 In this section, to compensate for the shooting method presented in the main text, we present an alternative numerical method that constructs an approximate solution of the EL equation
 for the DM model [Eqs.~\eqref{eq:ELDM1}-\eqref{eq:ELDM4}].
 In doing so, we employ Physical Informed Neural Networks (PINNs) techniques for approximating the solution \cite{RAISSI2019} by minimizing the OM action subject to the boundary condition,
 $\bm{x}(0)=\bm{X}_{\rm init}$ and $\bm{x}(\tau)=\bm{X}_{\rm final}$.

To numerically approximate the path $\{\bm{x}\}_{t=0}^{\tau}$,
we consider a function $\tilde{\bm{\phi}}_{\bm{\theta}}(t):  \mathbb{R} \to \mathbb{R}^4$ with parameters 
$\bm{\theta}$ based on Multi Layer Perceptron (MLP) \cite{rumelhart_learning_1986,hiruta2025ae}.
MLP is implemented by layers of biased linear maps and we used Gaussian Error Linear Unit (GELU) for the activation function.
We inserted three hidden layers between the input and output layers,
and the numbers of nodes are $16$, $128$, and $16$.

The boundary conditions 
are exactly satisfied by the following transformation,
\begin{align}
\bm{\phi}_{\bm{\theta}}(t) &= \sin\left(\frac{\pi t}{\tau}\right)\tilde{\bm{\phi}}_{\bm{\theta}}(t)+C\tilde{\bm{\phi}} +\bm{a} t +\bm{b},
\end{align}
with constant vectors $\bm{a}$ and $\bm{b}$ and a constant matrix $C$.
For the EL equations of the DM model, these constants are obtained as:
\begin{align}
\bm{a}&= \left(\frac{X_{\rm final}-X_{\rm init}}{\tau} ,\frac{Y_{\rm final}-Y_{\rm init}}{\tau},0,0\right),\\
\bm{b}&= (X_{\rm init} ,Y_{\rm init},0,0),\\
C&=\mathrm{diag} (0,0,1,1).
\end{align}
Here, $\bm{\phi}_{\theta}$ satisfies the boundary condition on the $x$ and $y$ components, whereas no specific boundary conditions are imposed on the $p_x$ and $p_y$ components.
The part
$\bm{a}t+\bm{b}$ corresponds to a constant momentum solution in the main text [Eq.~\eqref{eq:constmomentum}].

We introduce a loss function based on the ensemble average of the residual of the EL equations, $\langle R\rangle$, defined as
\begin{align}
    R(\bm{\phi},t)&\equiv \bigg| \frac{d\bm{\phi}_{\bm{\theta}}}{dt}(t)-\bm{f}(\bm{\phi}_{\bm{\theta}}(t))\bigg|^2
\end{align}
The time derivative $d\bm{\phi}_{\bm{\theta}} /dt$ is evaluated exactly and swiftly by automatic differentiation,
implemented via the machine-learning library PyTorch \cite{pytorch}.
Moreover, we introduce another loss function based on the action value $L_{\mathrm{action}}=S(\{\bm{\phi}_{\bm{\theta}}\})$ and evaluate the action value using the trapezoidal rule. For an illustrative example shown in Fig.~\ref{fig:posit_prob} with $\tau=5$, we used $128$ equally spaced time.

The total loss $L$ is then defined as the combination of $L_1$ and $L_2$,
\begin{align}
    L&=cL_{\mathrm{res}}+(1-c)L_{\mathrm{action}},
\end{align}
with a parameter $c\in[0,1]$.
The minimization problem of $L$ is solved using the Adam method and the learning rate is scheduled through a cosine annealing scheduler.

During the optimization procedure, we schedule for $c$ by
starting with $c=0$  and gradually increasing $c$ to $c=1$. 
The solutions and their actions well approximate the result from the shooting method as shown in Fig.~\ref{fig:posit_probA} with $\tau=5$ and $D=10^{-1}$ for different initial states, $\bm{X}_{\rm init}=\bm{Z}(\alpha)$. The nearly same result can also be obtained from different schedules such as fixing $c$ to one or zero throughout the optimization process.

\bibliography{bib}

\end{document}